%% file: O_stars_Cat.tex
\documentclass{aa}
\usepackage{graphicx}
\usepackage{txfonts}
\usepackage{natbib}
\bibpunct{(}{)}{;}{a}{}{,}

\def\mathstacksym#1#2#3#4#5{\def#1{\mathrel{\hbox to 0pt{\lower#5\hbox{#3}\hss} \raise #4\hbox{#2}}}}
\mathstacksym\gta{$>$}{$\sim$}{1.5pt}{3.5pt} 
\mathstacksym\lta{$<$}{$\sim$}{1.5pt}{3.5pt} 
\input{rrmacros}
\begin{document}
\title{The Origin of Massive O-type Field Stars}
\subtitle{Part I: A Search for Clusters\thanks{Based on observations obtained at the ESO-NTT,
La Silla, and at the Italian Telescopio Nazionale Galileo (TNG) operated on the island
of La Palma by the Centro Galileo Galilei of the CNAA
(Consorzio Nazionale per l'Astronomia e l'Astrofisica) at the
Spanish Observatorio del Roque de los Muchachos of the
Instituto de Astrofisica de Canarias.}}

\titlerunning{Massive Field Stars}
\author{        W.J. de Wit\inst{1} \and
                L. Testi\inst{1} \and
                F. Palla\inst{1} \and
                L. Vanzi\inst{2} \and
		H. Zinnecker\inst{3} 
                }
\offprints{W.J. de Wit, \email{dewit@arcetri.astro.it}}

\institute{     INAF, Osservatorio Astrofisico di Arcetri, Largo E. Fermi 5,
                50125 Florence, Italy \and
		ESO, Alonso de Cordova 3107, Santiago, Chile \and
		Astrophysikalisches Institut Potsdam, An der Sternwarte 16,
                14482 Potsdam, Germany}

\date{Received date; accepted date} 
\abstract{ We present a study aimed at
clarifying the birthplace for 43 massive O-type {\it field} stars. In this first paper
we present the observational part: a search for stellar clusters near the target stars.
We derive stellar density maps at two different resolving scales, viz. $\rm \sim
0.25\,pc~and~ \sim1.0\,pc$ from NTT and TNG imaging and the 2MASS catalogue.  These
scales are typical for cluster sizes. The main result is that the large majority of the
O-type field population are isolated stars: only 12\% (5 out of 43) of the O-type field
stars is found to harbour a small-scale stellar cluster.  We review the literature and
aim at characterizing the stellar field of each O-type field star with the emphasis on
star formation and the presence of known young stellar clusters. An analysis of the
result of this paper and a discussion of the O-type field population as products of a
dynamical ejection event is presented in an accompanying paper.  
} 
\maketitle

\section{Introduction}
We present the first of two papers that treat the search for the
origin of formation of massive O-type field stars in the Galaxy.
The O-type field stars are isolated massive stars in the Galactic
field and are not known to be members of stellar clusters and/or
OB-associations. These stars constitute about $20\%$ of the total number of 
O-type stars in the Galaxy (see e.g. Gies 1987\nocite{1987ApJS...64..545G}; Mason et al.
1998\nocite{1998AJ....115..821M}). Their relative small number and generally large
distances combined with an incomplete census of clusters/OB
associations throughout the Galaxy have led people to speculate
that the birthplace of the O-type field stars could
well be a cluster environment. If confirmed, this would be an
important finding in light of the ongoing discussion on the
formation of a high mass star, be it by stellar coalescence 
(Bonnell, Bate \& Zinnecker 1998\nocite{1998MNRAS.298...93B}; Bonnel, Vine \& Bate
2004\nocite{0401059}) or conversely by mass accretion through a
circumstellar disk/envelope (e.g. Yorke \& Sonnhalter
2002\nocite{ 2002ApJ...569..846Y}).

In this first contribution of our study on O-type field stars, we
present deep near infrared photometric observations to search for
clusters of low mass stars near the massive field stars.
Practically we probe for stellar density enhancements with respect to the
back/fore-ground stars on small ($\sim 5$~pc) and large ($\sim 20$~pc) scale. 
We complement the cluster search with a
literature overview for each target star in our sample with the
emphasis on the presence of nearby ($<65$\,pc) young stellar
clusters, associations or star forming regions. In this way we
hope to elucidate the formation history of massive field stars.
The discussion of the results of this study will be combined in an
accompanying paper with the interpretation of the massive O-type
field stars as the product of a dynamical ejection.


The organization of this paper is as follows. In Sect.\,\ref{number} we discuss
our working definition for field stars, the number of O-type stars in the field and
the sample selection. In Sect.\,\ref{obser} the observations and the analysis
method are described. Each O-type field star is reviewed and its stellar density
maps presented in Sect.\,\ref{fields}. We summarize the results in
Sect.\,\ref{summar}. 

\section{The number of O-type field stars}
\label{number}
\subsection{Definition of names}
\label{defin}
The definition of a field star is a negative one.
Field stars are those objects that are not member of any known
spatial concentration of stars, either a cluster or an association
of stars (and membership can be established using different
techniques). In general, field stars form the old population of a
Galaxy and should therefore be of low mass.


However, there exists a number of stars among the field population that are
known to be massive and therefore are much younger than the average age of the
field.  They come in two varieties. Historically, most attention was devoted to
those massive field stars that are known to have (or have had) high spatial
velocities, i.e. the runaway stars (see Blaauw 1961\nocite{1961BAN....15..265B}). From a stellar
population point of view these stars are a subset of the field, however when
deriving physical quantities of stellar clusters, runaways should be considered
as cluster members.

The second group of massive stars that are located in the field are simply those
O-type stars without large spatial velocities. They have received less attention
in the literature as a group, since it is expected that they are unrecognized
runaway stars (with yet to determine accurate proper motions) or that they are
part of an unrecognized cluster or star forming region. The primary cause for
the unknown origin of this subgroup of the O stars is that they are generally
located far away.

For clarity, in this paper we will make the explicit distinction between field 
O stars and runaway O stars, in the sense that runaway stars are not considered
to be part of the group of field O stars, following the above reasoning.

\subsection{Catalogues of O stars}

\begin{table}[t]
 {
  \begin{center}
 \caption[]{O star catalogues. In Col.\,1 the name and reference of the
  catalogue, Col.\,2 the total number of entries, Col.\,3 all entries with a
  apparent visual magnitude less than $\rm 8^{m}$. In Col.\,4 the subset and
  percentage of field stars with $\rm V<8^{m}$.}
  \begin{tabular}{lccc}
\hline
\hline
Name & $\rm N_{tot}$ & $\rm N_{tot}(V<8^{m})$ & $\rm N_{field}(V<8^{m})$ \\ 
\hline
COS82$^1$  & 765 &  205 & 91 (44\%) \\ 
G87$^2$  & 195 &  195 & 43 (22\%)   \\ 
M98$^3$  & 227 &  193 & 39 (20\%)   \\ 
M04$^4$  & 370 &  185 & 35 (19\%)   \\ 
\hline
\multicolumn{4}{l}{\tiny 1. Garmany, Conti \& Chiosi (1982); 2. Gies
  (1987); 3. Mason et al. (1998);}\\
\multicolumn{4}{l}{\tiny 4. Ma{\'{\i}}z-Apell{\'a}niz \& Walborn (2003)\nocite{2003IAUS..212..560M}}\\
 \end{tabular}
\label{tab_o}
\end{center}
}
\end{table}

\begin{figure}[t]
\includegraphics[height=8cm,width=8cm]{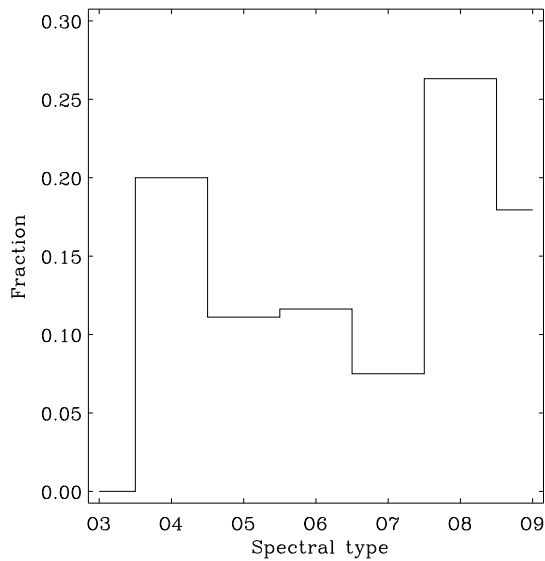}
  \caption[]{The fraction of field O stars to the total number of O-type stars 
  as function of spectral subtype, adopted from the M98 sample. Considering the
  small numbers, the fraction of field O stars seems independent of spectral subtype.}
\label{frac}
\end{figure}

Various catalogues of O stars have been compiled over the years.  The main
O-type star reference catalogue is the ``Catalog of O type Stars'', hereafter 
COS82, compiled from literature data by Garmany, Conti, \& Chiosi
(1982\nocite{1982ApJ...263..777G}). It is a volume limited sample out to an
estimated distance of $\rm \sim2.5\,kpc$ and contains 765
entries. Table\,\ref{tab_o} gives some characteristics of the COS82 catalogue
and of three other, more recent catalogues. These three catalogues are partly
based on the COS82 catalogue. All four catalogues list for every entry the
parent association/cluster when known. Using this information, Col.\,4 of
Table\,\ref{tab_o} shows the variation of the census of field O stars since
1982. It is reduced greatly by G87 with respect to COS82, and remains quite
stable at a fraction of $\sim20\%$ thereafter. The massive field star fraction
from the M98 catalogue is given as function of spectral subtype in
Fig.\,\ref{frac}. This fraction does not show a systematic dependence on spectral
type, which implies that if isolated massive star formation in the Galaxy
existed, the corresponding IMF would be similar to the one in clusters.  Small
differences in total number of entries between the G87, M98 and M04 catalogues
consist in O stars for which a clusters/association membership was established
over the years and for instance in the inclusion of O-type stars with Wolf-Rayet
secondaries (not included in M04, while 5 such systems are included in
M98). Finally, we mention that the difference between field O stars and runaway
O stars as laid down in the previous paragraph is not the common definition used
in these 4 catalogues. Therefore the field star numbers in Col\,4 of
Table\,\ref{tab_o} may include runaway stars.

In Table\,\ref{tab_o} we have not included the catalogue of Cruz-Gonz\'{a}lez et
al.  (1974)\nocite{1974RMxAA...1..211C}. Although the catalogue gives many
characteristics of its 664 entries, the catalogue does not specify memberships
since it primarily focuses on H~{\sc ii} regions and the ionization of the
interstellar medium.

\subsection{Sample selection}
\label{sample}
\begin{table*}[t]
\begin{minipage}{\textwidth}
 {
  \begin{center}
 \caption[]{Observations and specification of the O-type field stars as listed
  in M98. We adopted the spectral types (Col.\,2), distance (Col.\,4) and
  spectroscopic status (Col.\,5) from M98. The V magnitudes (Col.\,3) are from
  the GOS\footnote{http://www-int.stsci.edu/\hspace{0.1cm}$\tilde{ }$\hspace{0.1cm}jmaiz/GOSmain.html}
  compilation of Ma{\'{\i}}z-Apell{\'a}niz \& Walborn (2003).  Col.\,6
  gives the visual multiplicity status. The
  following abbreviations are used in Col.\,5: C = constant radial velocity, SB1
  = single-lined spectroscopic binary, SB2 = double-line spectroscopic
  binary. Addition of an "O" indicates that the orbit is known, an "E" indicates
  eclipsing systems. A column indicates uncertainty. In Col.\,6 we use OPT =
  optical binary, VB = visual binary, VMS = visual multiple object. In Col.\,9
  we give the K-band completeness limits for each observation and the
  corresponding mass limit (in $\rm M_{\odot}$). Col.\,11 gives measured 2MASS completeness limits.
  An asterisk at the O-type star's designation indicates a confirmed runaway status.
   }
  \begin{tabular}{llcrll|ccccc}
\hline
\hline
Name &       Sp. type &    V &    D & Spect.  &  Vis. & Obs. Date & Telescope &  $\rm K_{lim}$ & $\rm M_{lim}$&$\rm K_{lim}^{2MASS}$ \\
(1)  &         (2)   &   (3)&   (4)& (5)    &  (6) & (7) &    (8)    &   (9)   &  (10) & (11)       \\
\hline
\object{HD\,1337}   & O9.5III       & 5.90 & 2100 & SB2OE & single   & 2001-07-10 & TNG &  18.0  &$<0.10$& 15.0\\
\object{HD\,15137}  & O9.5II-III(n) & 7.87 & 3400 & SB2:  & single   & 2001-11-21 & TNG &  18.0  &0.15& 15.0\\
\object{HD\,36879}  & O7V(n)        & 7.57 & 1600 & C     & single   & 2001-02-02 & NTT &  17.0  &$<0.10$& 15.0\\
\object{HD\,39680}  & O6V(n)pevar   & 7.94 & 2500 & C     & OPT      & 2001-02-02 & NTT &  17.0  &$<0.10$& 15.0\\
\object{HD\,41161}  & O8Vn          & 6.77 & 1500 & C     & VB       & 2001-11-21 & TNG &  18.0  &$<0.10$& 15.0\\
\object{HD\,48279}  & O8V            & 7.91 & 2000 & C     & OPT      & 2001-02-02 & NTT & 17.5  &$<0.10$& 15.0 \\
\object{HD\,52266}  & O9IV(n)        & 7.21 & 1700 & SB1:  & single   & 2001-02-02 & NTT & 17.5  &$<0.10$& 15.0 \\
\object{HD\,52533}  & O8.5V          & 7.70 & 2000 & SB1O  & VMS      & 2001-02-02 & NTT & 17.0  &$<0.10$& 15.0\\
\object{HD\,57682}  & O9IV           & 6.42 & 1600 & C     & single   & 2001-02-02 & NTT & 17.0  &$<0.10$& 15.0\\\
\object{HD\,60848}  & O8Vpe          & 6.85 & 1900 & U     & single   & 2001-03-22 & NTT & 16.5  &0.14& 15.0\\
\object{HD\,66811}$^{*}$ & O4I(n)f    & 2.26 &  450 & C     & single   & 2001-02-02 & NTT & 17.0 &$<0.10$& 15.0 \\
\object{HD\,75222}$^{*}$ & O9.7Iab    & 7.42 & 2400 & C     & single   & 2001-02-02 & NTT & 17.0 &0.17& 15.0 \\
\object{HD\,89137}  & O9.5IIInp      & 7.98 & 3000 & SB1:  & single   & 2001-02-02 & NTT & 17.5  &0.18& 15.0\\
\object{HD\,91452}  & O9.5Iab-Ib     & 7.50 & 3100 & C     & single   & 2000-05-26 & NTT & 16.5  &0.39& 15.0\\
\object{HD\,96917}  & O8.5Ib(f)      & 7.08 & 2700 & SB1:  & single   & 2000-05-25 & NTT & 16.5  &0.30& 15.0\\
\object{HD\,105627} & O9II-III       & 8.14 & 3200 & C     & OPT      & 2000-05-25 & NTT & 15.5  &0.82& 15.0\\
\object{HD\,112244} & O8.5Iab(f)     & 5.38 & 1500 & SB1:  & VB       & 2000-05-25 & NTT & 16.7  &$<0.10$& 15.0\\
\object{HD\,113659} & O8/9III        & 8.05 & 1700 & U     & single   & 2001-03-22 & NTT & 16.5  &$<0.10$& 14.0\\
\object{HD\,117856} & O9.5III        & 7.38 & 1700 & SB2:  & VB       & 2000-05-26 & NTT & 16.5  &0.15& 14.0\\
\object{HD\,120678} & O8IIInep       & 7.87 & 2200 & U     & single   & 2001-02-02 & NTT & 16.5  &0.21& 13.5\\
\object{HD\,122879} & O9.5I          & 6.43 & 2100 & C     & single   & 2001-02-02 & NTT & 17.0  &0.15& 13.5\\
\object{HD\,123056} & O9.5V((n))     & 8.14 & 1600 & C     & single   & 2000-05-25 & NTT & 16.5  &$<0.10$& 14.0\\
\object{HD\,124314} & O6V(n)((f))    & 6.64 & 1000 & SB1:  & VB       & 2000-05-25 & NTT & 15.5  &$<0.10$& 14.0\\
\object{HD\,125206} & O9.5IV(n)      & 7.92 & 1700 & SB2:  & single   & 2001-03-22 & NTT & 15.5  &0.33& 14.0\\
\object{HD\,135240} & O7.5III((f))   & 5.08 &  920 & SB2OE & OPT      & 2001-03-22 & NTT & 16.5  &$<0.10$& 14.0\\
\object{HD\,135591} & O7.5III((f))   & 5.46 & 1100 & C     & VMS      & 2001-03-22 & NTT & 16.5  &$<0.10$& 14.0\\
\object{HD\,153426} & O9II-III       & 7.47 & 2100 & SB2:  & single   & 2000-05-25 & NTT & 16.2  &0.25& 14.0\\
\object{HD\,153919}$^{*}$ & O6.5Iaf   & 6.55 & 1700 & SB1OE & single   & 2001-03-22 & NTT & 16.5 &$<0.10$& 14.0 \\
\object{HD\,154368} & O9.5Iab        & 6.13 & 1100 & SBE   & VB       & 2000-05-25 & NTT & 16.5  &$<0.10$& 14.0 \\
\object{HD\,154643} & O9.5III        & 7.17 & 1300 & SB1:  & single   & 2001-03-23 & NTT & 16.5  &$<0.10$& 13.5 \\
\object{HD\,154811} & OC9.7Iab       & 6.92 & 1800 & C     & single   & 2001-03-23 & NTT & 17.0  &$<0.10$& 14.5\\
\object{HD\,158186} & O9.5V          & 7.00 & 1100 & SBE   & single   & 2001-03-23 & NTT & 13.0  &0.82& 12.5\\
\object{HD\,161853} & O8V((n))       & 7.92 & 1600 & SB1:  & single   & 2001-03-23 & NTT & 13.5  &0.91&12.0\\
\object{HD\,163758} & O6.5Iaf        & 7.33 & 3600 & C     & single   & 2001-03-23 & NTT & 16.5  &0.38&13.0\\
\object{HD\,165319} & O9.5Iab        & 7.93 & 2100 & C     & single   & 2000-05-25 & NTT & 14.0  &1.27&13.0\\
\object{HD\,169515} & O9.7Ibpevar    & 9.19 & 2000 & SB2OE & single   & 2000-05-25 & NTT & 14.0  &1.10& 12.5\\
\object{HD\,175754} & O8II((f))      & 7.02 & 2700 & C     & single   & 2000-05-25 & NTT & 17.0  &0.19& 15.0\\
\object{HD\,175876} & O6.5III(n)     & 6.94 & 2300 & C     & OPT      & 2000-05-25 & NTT & 16.0  &0.29& 15.0\\
\object{HD\,188209} & O9.5Iab        & 5.63 & 2000 & C     & single   & 2001-07-07 & TNG & 18.4  &$<0.10$& 15.0\\
\object{HD\,193793} & O5             & 6.88 & 1300 & SB2O  & VB       & 2001-07-07 & TNG & 18.4  &$<0.10$& 15.0\\
\object{HD\,195592} & O9.7Ia         & 7.08 & 1400 & SB1:  & single   & 2001-07-06 & TNG & 18.0  &$<0.10$& 15.0\\
\object{HD\,201345} & ON9V           & 7.66 & 1900 & C     & single   & 2001-07-07 & TNG & 18.4  &$<0.10$& 15.0\\
\object{HD\,202124} & O9.5Iab        & 7.81 & 3500 & C     & single   & 2001-07-07 & TNG & 18.4  &$<0.10$& 15.0\\
\hline
\end{tabular}
\label{cat}
\end{center}
}
\end{minipage}
\end{table*}

In the present study we select the 43 field O stars from the M98 study, that
contain 39 stars with $\rm V <8^{m}$ (see Table\,\ref{tab_o}), plus 4 additional fainter objects.
These authors search for nearby ($0.035''<\rho<1.5''$) companion stars to the
target objects using speckle interferometry. They reach the following
conclusions with regard to the field O stars: at least 35\% contains a
secondary object (spectroscopic or visual), and 2 objects are visually
multiple systems. Table\,\ref{cat} specifies spectral type, visual magnitude,
distance, spectroscopic and visual multiplicity status for each of the field O
star in M98, sorted on HD number. The following abbreviations are used for the
spectroscopic status: C = constant radial velocity (20 stars), SB1 = single-line
spectroscopic binary (8 stars) , SB2 = double-line spectroscopic binary (7 stars). Addition
of the letter ``O'' indicates that the orbit is known (6 stars), an ``E'' indicates
eclipsing systems (6 stars). A colon has the usual meaning of uncertainty. For the
visual multiplicity status in Col.\,6, following G87, we use OPT = optical
binary (5 stars), VB = visual binary (6 stars), VMS = visual multiple object (2 stars) and
single O-type field stars (30 stars).

The M98 study is an update of the study by G87. Therefore the M98 census of the
43 field O stars nearly equals the G87 census. For completeness reasons we
point out that the M98 sample differs by four objects with G87, that were in fact
discovered to be member of a cluster or association, viz. HD\,5005, HD\,14947,
HD\,47432, HD\,71304.  

We now give a short description of how the O star field sample is
compiled with respect to identification of parent clusters. The O-type star
sample of G87 is constructed primarily from the LSC catalogue (Humphreys \&
McElroy 1984), selecting stars with $V<8^{m}$. In turn, the main O-type star
input of the LSC is the COS82. The cluster/association membership for the stars
is obtained by cross-correlating the LSC\nocite{1984ApJ...284..565H} with the
clusters and associations as catalogued by Alter, Ruprecht \& Van\'{y}sek
(1970)\nocite{1970CSCA..C......0A} and Ruprecht, Bal\'{a}zs \& White
(1981)\nocite{1981csca.book.....R}. The {\it runaway} O stars in the M98 sample
are defined using one of three separate criteria: (1) having either a peculiar
radial velocity greater than 30\,$\rm km\,s^{-1}$, (2) a peculiar space velocity
greater than 30\,$\rm km\,s^{-1}$, or (3) stars with a distance from the
Galactic plane greater than 500\,pc. Finally, 43 O-type stars in the M98 sample
are not members of a cluster and are not runaway stars. They are annotated as {\it
field} stars. We would like to note that the field O star population defined in
this way contains three confirmed runaway stars, viz. HD\,66811 ($\zeta$~Pup,
e.g.  Chlebowski 1989\nocite{1989ApJ...342.1091C}), HD\,75222 (Hoogerwerf et
al. 2001\nocite{2001A&A...365...49H}) and HD\,153919 (Ankay et
al. 2001\nocite{2001A&A...370..170A}). We have marked them in Table\,\ref{cat}
with an asterisk.

As a final note on the M98 field O star census, we would like
to point out the existing confusion concerning the visual binary
star HD\,48149. The
star is characterized in the M98 sample as a field O8.5V star. However, according to
SIMBAD, it is a G2/G3V star. SIMBAD gives the extra note that
confusion exists between this star and \object{HD\,46149}, an 08.5V star in
the cluster \object{NGC\,2244}. We do not consider HD\,48149 as an O-type star
in this paper.

\section{Observations and search method}
\label{obser}
\subsection{ESO/NTT, INAF/TNG and 2MASS}
The 43 field O stars from the M98 census were observed with the NTT at the ESO
LaSilla facility and with the Telescopio Nazionale Galileo (TNG). The TNG
is Italy's 3.5m optical/infrared telescope, operated from the Roque de los Muchachos on La
Palma (Canary Islands). It has an alt-azimuth mount and
Ritchey-Chretien configuration with two Nasmyth foci and active optics control.

\begin{figure}[t]
\includegraphics[height=8cm,width=8cm]{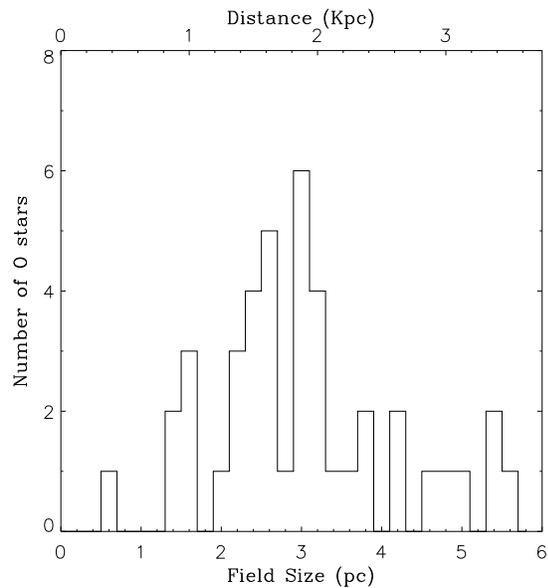}
  \caption[]{Distribution of distances of target O-type stars and the
  corresponding sizes of the field of view.} 
 \label{fov}
\end{figure}

At the NTT the stars were observed with the SOFI instrument, whereas the TNG was
equipped with NICS (Near Infrared Camera Spectrometer) camera. The field
of view (FoV) and resolution using these cameras are very similar, $5\times 5$
arcmin and $\sim 0.3$~arcsec/pixel, respectively. In Fig.\,\ref{fov} we show the
distribution of the projected linear scale of the FoV of SOFI and NICS for the
adopted distances listed in Table\,\ref{cat}.  NTT and 2MASS imaging was
performed in the NIR K-short filter ($\rm \lambda_{c}=2.16\mu m$) and TNG imaging in the K-prime
 filter ($\rm \lambda_{c}=2.12\mu m$) , using standard dithering strategy
for background subtraction purposes. The K-band was chosen to minimize the
difference in magnitude between the bright O-type star and the surrounding
stellar population. Images were bias-subtracted, flat-fielded and combined using
NOAO/IRAF data reduction software. Source extraction and aperture photometry was
done with the DAOPHOT routines running under IDL.

The central O-type stars are in all cases saturated, because long
exposure times were used to pick up the faintest stars. The
saturated profile is highly non-linear and shows features that are
identified by DAOPHOT as candidate stars. Meticulous care is taken
in source identification close to the saturated target O-type
stars. To prevent false detections, we compare different saturated
PSFs of the central O stars and other bright stars in the field to
identify the general PSF features. Such exercises result generally in
rejection of features dimmer than $\rm \Delta K \gta 8^{m}$ within a
radius of 5 arcsec from the target O star.

To probe larger fields in search for clusters  around
field O stars, we use the recently released 2MASS All Sky Data
Release. Catalogues spanning $20 \times 20$ arcmin were
extracted from the Point Source Catalog (PSC)\footnote{from
http://www.ipac.caltech.edu/2mass/} and analyzed in the same way
as the NTT and TNG data. The PSC is 99 \% complete for $\rm K_{s}<
14.3^{m}$, however we use for each 2MASS field the proper completeness
limit (see Col.\,10 of Table\,\ref{cat}).

\subsection{Counting method}
The images are analyzed for stellar density enhancements, by
counting and binning methods. In each field the detected
stars are binned in subfields. For the deeper TNG and NTT images,
we choose a binning resolution of $11 \times 11$ subfields. The shallower
but larger 2MASS fields are all subdivided in $15 \times 15$ subfields. The
binning grid is chosen so that the target O-type star falls
in the centre of one of the subfields. Then, the average stellar
density is calculated and the corresponding Poisson standard deviation compared
with the number of detected stars in each subfield. A 3\hspace{0.05cm}$\sigma$
offset from the average is considered to be a possible
cluster of stars, although it could still be compatible with the
Poisson distribution of the number of stars per subfield. Whenever
the 3\hspace{0.05cm}$\sigma$ detection also happens to be centred on the target
star, its physical association is inferred. 

Only stars brighter than the completeness limit are taken into
account. The completeness limits (Cols.\,9 and 10 in
Table\,\ref{cat}) differ from frame to frame due to crowding and observing conditions. The
exposure time is the same for each field. The density maps are
presented in Sect.\,\ref{fields} with overlaid contours indicating
the offset in standard deviations, starting at 1\hspace{0.05cm}$\sigma$ from the mean density of the complete
field. We note that the density maps have different
linear resolutions, because the binning is chosen the same for
each field. Therefore 2MASS maps have linear resolutions between 0.5\,pc
and 2\,pc, whereas the TNG and NTT maps between 0.15\,pc and
0.5\,pc. The linear resolutions of the density maps are
comparable to the sizes of stellar clusters. The maps are smoothed
for presentation purposes. Some maps show white, low-density
regions near bright, saturated stars (e.g. HD\,91452 or
HD\,125206) and also near some of the target stars. This is caused
by the extended and highly non-linear PSF of saturated objects in
the SOFI and NICS images.

\section{O-type field stars; visually single or in clusters?}
\label{fields}

In this section we present the K-band images, the stellar densities maps on
small (NTT and TNG) and large scales (2MASS), and a description of
each object. The description focuses on the currently known visual multiplicity
status of the field O stars and nearby early-type ($\rm < B\,5$) stars. Special
attention is given to nearby young clusters
taken from the literature and/or from
WEBDA\footnote{http://obswww.unige.ch/webda/navigation.html}. 
Specifically, we searched for known stellar
clusters with ages less than $\rm 10\,Myr$ located within a projected distance
less than 65\,pc as calculated from the distance of the field O star. The
average peculiar radial velocity of field O stars is $\sim \rm 6.4 \,km\,s^{-1}$ (G87), therefore
during a lifetime of $\sim10^7$\,years, the field O stars may wander about $\rm
\sim 65\,pc$ from their birthplace. In the case
that such a cluster exists, and its distance from the
Sun is similar to the field O stars' distance within the uncertainty, 
we list its basic characteristics in a ``mini table'' added to the
description of that particular star. We adopting an uncertainty of
30\% on the distances for the O stars.  The stars are presented in this
section ordered by their HD number.

Beforehand we'd like to note that in nearly all cases, the 2MASS maps
3\hspace{0.05cm}$\sigma$ ``detection'' show no other indications for a physical
cluster, e.g. bright stars or nebulosities either from DSS images or after
cross-correlation with the SIMBAD database. These detections are probably purely
statistical.  The two exceptions are HD\,48279 (Fig.\,\ref{f_hd4827} a new 2MASS
cluster detected on the edge of the considered field) and HD\,52533
(Fig.\,\ref{f_hd5768}, a 2MASS cluster centred on the target star). For every
object in the following list, we classify a density peak by stating whether it
is a {\it physical} as opposed to {\it statistical} cluster depending on whether
it is associated with the above mentioned secondary indicators.


\input{FIG_TEX/Fig3AP.tex}

\input{STARS_TEX/hd1337.tex}
\input{STARS_TEX/hd15137.tex}

\input{STARS_TEX/hd36879.tex}

\input{FIG_TEX/Fig4AP.tex}

\input{STARS_TEX/hd39680.tex}

\input{STARS_TEX/hd41161.tex}
\input{STARS_TEX/hd48279.tex}

\input{FIG_TEX/Fig5AP.tex}

\input{STARS_TEX/hd52266.tex}

\input{STARS_TEX/hd52533.tex}
\input{STARS_TEX/hd57682.tex}

\input{FIG_TEX/Fig6AP.tex}

\input{STARS_TEX/hd60848.tex}
\input{STARS_TEX/hd66811.tex}
\input{STARS_TEX/hd75222.tex}

\input{FIG_TEX/Fig7AP.tex}

\input{STARS_TEX/hd89137.tex}
\input{STARS_TEX/hd91452.tex}
\input{STARS_TEX/hd96917.tex}

\input{FIG_TEX/Fig8AP.tex}

\input{STARS_TEX/hd105627.tex}
\input{STARS_TEX/hd112244.tex}
\input{STARS_TEX/hd113659.tex}

\input{FIG_TEX/Fig9AP.tex}

\input{STARS_TEX/hd117856.tex}
\input{STARS_TEX/hd120678.tex}

\input{STARS_TEX/hd122879.tex}

\input{FIG_TEX/Fig10AP.tex}

\input{STARS_TEX/hd123056.tex}
\input{STARS_TEX/hd124314.tex}

\input{STARS_TEX/hd125206.tex}

\input{FIG_TEX/Fig11AP.tex}

\input{STARS_TEX/hd135240.tex}
\input{STARS_TEX/hd135591.tex}
\input{STARS_TEX/hd153426.tex}

\input{FIG_TEX/Fig12AP.tex}

\input{STARS_TEX/hd153919.tex}
\input{STARS_TEX/hd154368.tex}
\input{STARS_TEX/hd154643.tex}

\input{FIG_TEX/Fig13AP.tex}

\input{STARS_TEX/hd154811.tex}
\input{STARS_TEX/hd158186.tex}

\input{STARS_TEX/hd161853.tex}

\input{FIG_TEX/Fig14AP.tex}

\input{STARS_TEX/hd163758.tex}
\input{STARS_TEX/hd165319.tex}
\input{STARS_TEX/hd169515.tex}

\input{FIG_TEX/Fig15AP.tex}

\input{STARS_TEX/hd175754.tex}
\input{STARS_TEX/hd175876.tex}
\input{STARS_TEX/hd188209.tex}

\input{FIG_TEX/Fig16AP.tex}

\input{STARS_TEX/hd193793.tex}
\input{STARS_TEX/hd195592.tex}
\input{STARS_TEX/hd201345.tex}

\input{FIG_TEX/Fig17AP.tex}

\input{STARS_TEX/hd202124.tex}

\section{Summary}
\label{summar}

In this first contribution of two papers on massive O-type field stars, we have
presented and discussed stellar density maps derived from K-band
imaging. The maps were used to search for stellar clusters near 43 target field
stars in order to test the hypothesis whether massive field stars are actually
members of yet unrecognized clusters. The O-type field star sample is adopted
from the paper by Mason et al. (1998). Searches for relatively small scale
clusters ($\rm \sim0.25\,pc$) were executed using deep K-band
observations, generally with sub-solar mass completeness limits. Larger cluster
sizes ($\rm \sim1\,pc$) were probed using the 2MASS All Sky Data Release. In
general, the presented density maps show that the O-type field stars are not
residing in clusters. Only in a few instances the maps indicate the presence of
stellar density peaks near the target stars. The stars which we found in a
cluster are: HD\,52266, HD\,52533, HD\,57682, HD\,153426, and HD\,195592.
The best of these cases is for stars HD\,52533 and HD\,195592. 

The main result of this paper being the general absence of stellar clusters near
the massive field O stars is compatible with the suggestion made by Gies (1987). The
author lists a number of statistical arguments why the sample of O-type field
stars is not compatible with a formation origin in situ. The arguments are based
on the kinematical, binary and spectral type statistics of the O-type stars
compared to similar stars in clusters/associations and the O-type runaway stars.
According to Gies, the O-type field stars could constitute the low velocity
tail of the velocity distribution due to dynamical ejection from young stellar
clusters. To investigate this in more detail, the accompanying paper (de Wit et
al. 2004\nocite{PAPERII}) will extend the discussion in light of the main result of this paper
and use additional information on spatial velocities, association with nearby
young clusters and star formation regions in order to constrain the origin of the
massive O-type field stars.



\begin{acknowledgements} WJDW acknowledges financial assistance from the European Union Research
Training Network `The Formation and Evolution of Young Stellar Clusters'
(RTN1-1999-00436).  This research has made use of the Sim\-bad database,
operated at the Centre de Don\-nees Astro\-no\-miques de Stras\-bourg (CDS), and
of NASA's Astrophysics Data System Biblio\-graphic Services (ADS). This
publication makes use of data products from the Two Micron All Sky Survey, which
is a joint project of the University of Massa\-chusetts and the Infrared
Processing and Analysis Center/California Institute of Tech\-nology, funded by
the National Aeronautics and Space Admini\-stration and the National Science
Foundation.
\end{acknowledgements}

\bibliographystyle{aa}
\bibliography{/tmp/wdwfiles.bib}

\end{document}

%% file: rrmacros.tex
\newcount\longrefs
\def\aap{\ifnum\longrefs=1 {Astron.\ Astrophys.}\else 
                           {A\hbox{\rm \&}A}\fi}
\def\aapr{\ifnum\longrefs=1 {Astron.\ Astrophys.\ Rev.}\else 
                            {A\hbox{\rm \&}AR}\fi}
\def\aaps{\ifnum\longrefs=1 {Astron.\ Astrophys.\ Suppl.}\else 
                            {A\hbox{\rm \&}A Suppl.}\fi}
\def\aj{\ifnum\longrefs=1 {Astron.\ J.}\else 
                          {AJ}\fi} 
\def\ao{\ifnum\longrefs=1 {Applied Optics}\else 
                           {Appl.\ Opt.}\fi} 
\def\apj{\ifnum\longrefs=1 {Astrophys.\ J.}\else 
                           {ApJ}\fi} 
\def\apjl{\ifnum\longrefs=1 {Astrophys.\ J. Lett.}\else 
                            {ApJ}\fi} 
\def\aplett{\ifnum\longrefs=1 {Astrophys.\ J. Lett.}\else 
                            {ApJ}\fi} 
\def\apjs{\ifnum\longrefs=1 {Astrophys.\ J. Suppl.}\else 
                            {ApJS}\fi}
\def\apss{\ifnum\longrefs=1 {Astrophys.\ and Space Science}\else 
                            {Astrophys.\ Space Sci.}\fi}
\def\azh{\ifnum\longrefs=1 {Astronomicheskii Zhurnal}\else 
                            {Astron.\ Zhur.}\fi}
\def\araa{\ifnum\longrefs=1 {Ann.\ Rev.\ Astron.\ Astrophys.}\else 
                            {ARA\hbox{\rm \&}A}\fi}
\def\baas{\ifnum\longrefs=1 {Bull.\ Am.\ Astron.\ Soc.}\else 
                            {BAAS}\fi}
\def\bain{\ifnum\longrefs=1 {Bull.\ Astronom.\ Institutes Netherlands}\else
                            {Bull.\ Astr.\ Inst.\ Neth.}\fi}
\def\gca{\ifnum\longrefs=1 {Geochim.\ Cosmochim.\ Acta}\else 
                           {Geochim.\ Cosmochim.\ Acta}\fi}
\def\grl{\ifnum\longrefs=1 {Geophys.\ Res.\ Lett.}\else 
                           {Geoph.\ Res.\ Lett.}\fi}
\def\iaucirc{\ifnum\longrefs=1 {IAU Circulars}\else 
                          {IAU Circ.}\fi}
\def\ip{\ifnum\longrefs=1 {in press}\else 
                          {in press}\fi}
\def\jgr{\ifnum\longrefs=1 {J.\ Geophys.\ Res.}\else 
                           {J.\ Geophys.\ Res.}\fi}  
\def\jrasc{\ifnum\longrefs=1 {J.\ Royal Astron.\ Soc.\ Canada}\else 
                           {JRAS Can.}\fi}  
\def\mnras{\ifnum\longrefs=1 {Mon.\ Not.\ Roy.\ Astron.\ Soc.}\else 
                             {MNRAS}\fi} 
\def\memras{\ifnum\longrefs=1 {Mem.\ R.\ Astron.\ Soc.}\else 
                             {MmRAS}\fi} 
\def\nat{\ifnum\longrefs=1 {Nature}\else 
                           {Nat}\fi}
\def\pasj{\ifnum\longrefs=1 {Pub.\ Astron.\ Soc.\ Japan}\else 
                            {PASJ}\fi} 
\def\pasp{\ifnum\longrefs=1 {Pub.\ Astron.\ Soc.\ Pacific}\else 
                            {PASP}\fi} 
\def\physscr{\ifnum\longrefs=1 {Physica Scripta}\else 
                            {Phys.\ Scrip.}\fi} 
\def\planss{\ifnum\longrefs=1 {Planetary \& Space Science}\else 
                            {Plan. \& Space Sci.}\fi} 
\def\procspie{\ifnum\longrefs=1 {Proc.\ SPIE}\else 
                            {Proc.\ SPIE}\fi} 
\def\qjras{\ifnum\longrefs=1 {Quarterly J.\ Royal Astron.\ Soc.}\else 
                            {QJRAS}\fi} 
\def\skytel{\ifnum\longrefs=1 {Sky \& Telescope}\else 
                            {Sky \& Tel.}\fi} 
\def\solphys{\ifnum\longrefs=1 {Solar Phys.}\else 
                               {Solar Phys.}\fi}

\def\dutch{\def\refname{Referenties}\def\abstractname{Samenvatting}%
  \def\bibname{Bibliografie}\def\chaptername{Hoofdstuk}%
  \def\appendixname{Bijlage}\def\contentsname{Inhoudsopgave}%
  \def\listfigurename{Lijst van figuren}\def\listtablename{Lijst van tabellen}%
  \def\indexname{Index}\def\figurename{Figuur}\def\tablename{Tabel}%
  \def\partname{Deel}\def\enclname{Bijlage(n)}\def\ccname{Ter attentie van}%
  \def\headtoname{Aan}\def\headpagename{Pagina}%
  \def\today{\number\day\space\ifcase\month\or januari\or februari\or maart\or%
     april\or mei\or juni\or juli\or augustus\or september\or oktober\or%
     november\or december\fi \space\number\year}%
  \typeout{
              >>>>> use hlatex209 for Dutch hyphenation <<<<< 
         }}
\newcounter{onefig} \newcounter{fignumber}
\newcount\nocaptions \newcount\nofigures \newcount\figwidth
\newcount\viewgraphs
  \def\paper{}  \def\figlabel{} 
\long\def\nextfig#1{\setcounter{figure}{\value{fignumber}}
  \addtocounter{fignumber}{1}
  \ifnum \viewgraphs=1 \newpage \pagestyle{empty} \fi 
  \ifnum\value{onefig}=0 #1 \fi                 
  \ifnum\value{onefig}=\value{fignumber} #1 \fi}
\def\figwidths#1#2{\ifnum \nocaptions=1 #2mm \else #1mm \fi}  
\def\paper#1{}  
\long\def\plotfig#1#2{\ifnum \nofigures=1 \else #2 \fi}
\long\def\captiontext#1{\ifnum \nofigures=1 \raggedright \fi 
   \ifnum \nocaptions=1 \paper
     \ifnum \viewgraphs=0 
       \newline  \mbox{}\hrulefill\mbox{} \newline 
       \newline label:~\{\figlabel\} 
     \fi 
     \else \ifnum \nofigures=0 \fi 
   #1 \fi}

\newcount\panelwidth \newcount\panelheight 
\newcount\bxmin \newcount\bymin \newcount\bxmax \newcount\bymax
\newcount\tbxmin \newcount\tbymin
\newcount\tpanelwidth \newcount\tpanelheight \newcount\tpdif
\def\panelsize #1,#2;{\panelwidth=#1 \panelheight=#2}  
\def\setbb #1,#2;#3,#4;#5,#6;{
  \tbxmin=#1 \tbymin=#2    
  \bxmin=#3 \bymin=#4      
  \bxmax=#5 \bymax=#6}     
\def\barepanel #1{%
  \ifnum\panelheight=0 
    \tpdif=\bymax \advance\tpdif by -\bymin
    \multiply \tpdif by \panelwidth
    \tpanelheight=\tpdif
    \tpdif=\bxmax \advance\tpdif by -\bxmin
    \divide \tpanelheight by \tpdif
  \else \tpanelheight=\panelheight \fi
  \psfig{figure=#1,%
     bbllx=\bxmin bp,bblly=\bymin bp,bburx=\bxmax bp,bbury=\bymax bp,clip=t,%
     width=\panelwidth mm,height=\tpanelheight mm}}
\def\labelypanel #1{
  \ifnum\panelheight=0 
    \tpdif=\bymax \advance\tpdif by -\bymin
    \multiply \tpdif by \panelwidth
    \tpanelheight=\tpdif
    \tpdif=\bxmax \advance\tpdif by -\bxmin
    \divide \tpanelheight by \tpdif
  \else \tpanelheight=\panelheight \fi
  \tpdif=\bxmax \advance\tpdif by -\tbxmin
  \tpanelwidth=\panelwidth \multiply \tpanelwidth by \tpdif
  \tpdif=\bxmax \advance\tpdif by -\bxmin
  \divide \tpanelwidth by \tpdif
  \psfig{figure=#1,%
    bbllx=\tbxmin bp,bblly=\bymin bp,bburx=\bxmax bp,bbury=\bymax bp,%
    clip=t,width=\tpanelwidth mm,height=\tpanelheight mm}}
\def\labelxpanel #1{%
  \ifnum\panelheight=0 
    \tpdif=\bymax \advance\tpdif by -\bymin
    \multiply \tpdif by \panelwidth
    \tpanelheight=\tpdif
    \tpdif=\bxmax \advance\tpdif by -\bxmin
    \divide \tpanelheight by \tpdif
  \else \tpanelheight=\panelheight \fi
  \tpdif=\bymax \advance\tpdif by -\tbymin
  \multiply \tpanelheight by \tpdif
  \tpdif=\bymax \advance\tpdif by -\bymin
  \divide \tpanelheight by \tpdif
  \psfig{figure=#1,%
    bbllx=\bxmin bp,bblly=\tbymin bp,bburx=\bxmax bp,bbury=\bymax bp,%
    clip=t,width=\panelwidth mm,height=\tpanelheight mm}}
\def\labelxypanel #1{%
  \ifnum\panelheight=0 
    \tpdif=\bymax \advance\tpdif by -\bymin
    \multiply \tpdif by \panelwidth
    \tpanelheight=\tpdif
    \tpdif=\bxmax \advance\tpdif by -\bxmin
    \divide \tpanelheight by \tpdif
  \else \tpanelheight=\panelheight \fi
  \tpdif=\bxmax \advance\tpdif by -\tbxmin
  \tpanelwidth=\panelwidth \multiply \tpanelwidth by \tpdif
  \tpdif=\bxmax \advance\tpdif by -\bxmin
  \divide \tpanelwidth by \tpdif 
  \tpdif=\bymax \advance\tpdif by -\tbymin 
  \multiply \tpanelheight by \tpdif
  \tpdif=\bymax \advance\tpdif by -\bymin
  \divide \tpanelheight by \tpdif
  \psfig{figure=#1,%
    bbllx=\tbxmin bp,bblly=\tbymin bp,bburx=\bxmax bp,bbury=\bymax bp,%
    clip=t,width=\tpanelwidth mm,height=\tpanelheight mm}}

\def\CC{\par \vspace*{-2ex} \footnotesize \baselineskip=8pt \begin{verbatim}}

\long\def\startignore #1\stopignore{}   

\def\setlistparams{         
  \topsep=0.7ex                 
  \itemsep=0.7ex                
  \leftmargini=3ex}             
\setlistparams                  

\newcounter{alistindex}       

\newcounter{romenumnr}

\newlength{\minipagewidth}

\newsavebox{\boxcontent}
\newcommand{\ovalhead}[1]{
  \unitlength=1cm
  \sbox{\boxcontent}{\mbox{~~{#1}~~}}
  \begin{center}
    \ifdim\wd\boxcontent>6ex 
    \ifdim\wd\boxcontent<8cm 
    \begin{picture}(8,3) \thicklines     
      \put(4.0,0.8){\oval(8,1.6)} 
      \put(0.0,0.7){\parbox{8cm}{
         \begin{center} \usebox{\boxcontent} \end{center}}}
    \end{picture}
    \else \ifdim\wd\boxcontent<12cm 
    \begin{picture}(12,3) \thicklines     
        \put(6.0,0.8){\oval(12,1.6)} 
        \put(0.0,0.7){\parbox{12cm}{
           \begin{center} \usebox{\boxcontent} \end{center}}}
    \end{picture}
    \else
    \begin{picture}(16,3) \thicklines     
        \put(8.0,0.8){\oval(16,1.6)} 
        \put(0.0,0.7){\parbox{16cm}{
           \begin{center} \usebox{\boxcontent} \end{center}}}
    \end{picture}
    \fi \fi \fi
  \end{center}} 

\newcounter{headnr}            
\newcounter{subheadnr}[headnr]
\newcounter{subsubheadnr}[subheadnr]
\def\head #1\par{
  \stepcounter{headnr}                          
  \vspace{2ex} \noindent                        
  {\bf \theheadnr~~~~#1}\\[1ex] \noindent}      
\def\subhead #1\par{  
  \stepcounter{subheadnr}
  \vspace{1.3ex} \noindent
  {\bf \theheadnr.\arabic{subheadnr}~~~#1}\\[0.3ex] \noindent}
\def\subsubhead #1\par{
  \stepcounter{subsubheadnr}
  \vspace{1.0ex} \noindent
  {\bf \theheadnr.\arabic{subheadnr}.\arabic{subsubheadnr}~~~#1}\\ \noindent}

\font\dropfont= cmr12 scaled \magstep5
\def\dropcap#1#2{{\noindent
    \setbox0\hbox{\dropfont #1}\setbox1\hbox{#2}\setbox2\hbox{(}%
    \count0=\ht0\advance\count0 by\dp0\count1\baselineskip
    \advance\count0 by-\ht1\advance\count0by\ht2
    \dimen1=.5ex\advance\count0by\dimen1\divide\count0 by\count1
    \advance\count0 by1\dimen0\wd0
    \advance\dimen0 by.25em\dimen1=\ht0\advance\dimen1 by-\ht1
    \global\hangindent\dimen0\global\hangafter-\count0
    \hskip-\dimen0\setbox0\hbox to\dimen0{\raise-\dimen1\box0\hss}%
    \dp0=0in\ht0=0in\box0}#2}

\def\level #1 #2#3#4{$#1 \: ^{#2} \mbox{#3} ^{#4}$}   

\def\mathstacksym#1#2#3#4#5{\def#1{\mathrel{\hbox to 0pt{\lower 
    #5\hbox{#3}\hss} \raise #4\hbox{#2}}}}

\mathstacksym\lta{$<$}{$\sim$}{1.5pt}{3.5pt} 
\mathstacksym\gta{$>$}{$\sim$}{1.5pt}{3.5pt} 
\mathstacksym\lrarrow{$\leftarrow$}{$\rightarrow$}{2pt}{1pt} 
\mathstacksym\lessgreat{$>$}{$<$}{3pt}{3pt} 


%% file: FIG_TEX/Fig3AP.tex
\begin{figure*}[ht]
 \center
\includegraphics[height=15cm,width=16cm,angle=-90]{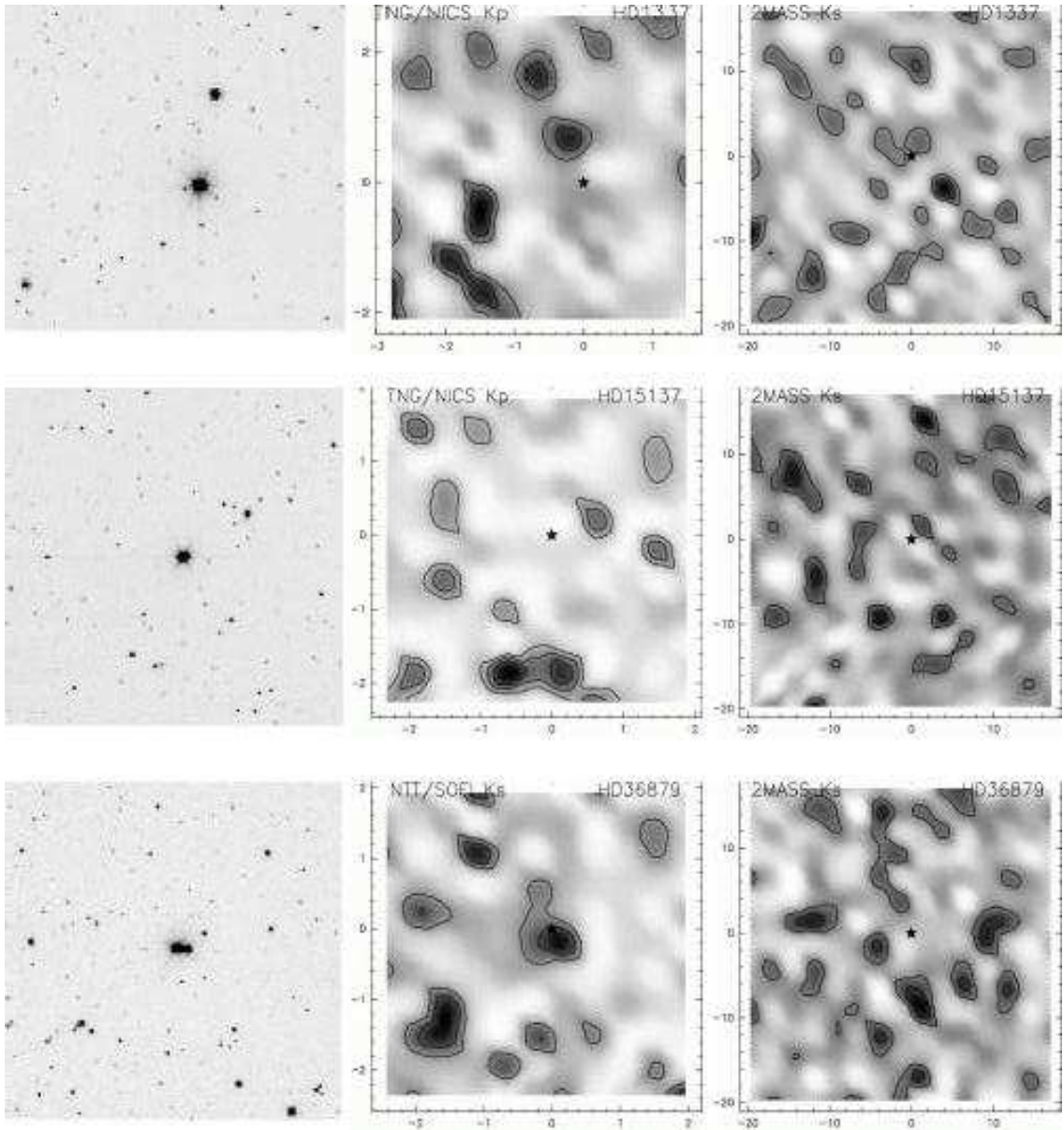}
 \caption[]{Left panels: $\rm K_{s}$-band image with 5 arcmin on a side.
 Middle panels: correspoding surface stellar density maps. Right panels:
 density maps derived from the 2MASS PSC, for stars HD\,1337, HD\,15137 and
 HD\,36879.  Ordinates are in arcmin for the middle and right panels. Contours are 1\vspace{0.05cm}$\sigma$ spaced deviations from
 the average stellar density value.}
\label{f_hd3687}
\end{figure*}

%% file: STARS_TEX/hd1337.tex
\subsection{HD\,1337 ($\equiv$AO\,CAS)}

The massive eclipsing binary AO\,Cas (O9.5\,III\,+\,O8\,V, Bagn\-uolo \& Gies
1991) \nocite{ 1991ApJ...376..266B} is a well studied system that shows the
signatures of colliding winds in UV spectra (e.g. Gies \& Wiggs
1991\nocite{1991ApJ...375..321G}). The system is associated with the ROSAT Faint
Source Catalogue source \object{1RXS J001747.4+512549} (Voges
2000)\nocite{2000yCat.9029....0V}. Uncertain spectroscopic distance
determinations place it between 1.2\,kpc (Stone
1978\nocite{1978AJ.....83..393S}), 2.1\,kpc (M98), and 3.9\,kpc
(Cruz-Gonz\'{a}lez et al. 1974\nocite{1974RMxAA...1..211C}). For the latter
distance scale, HD\,1337 would be 750\,pc below the Galactic plane. There are no
young clusters within a projected radius of 65\,pc for the adopted distance of
2.1\,kpc.

{\it Our result:} The stellar density maps in the middle and right panel of the top row in 
Fig.\,\ref{f_hd3687} do not indicate a cluster near HD\,1337. 

%% file: STARS_TEX/hd15137.tex
\subsection{HD\,15137}

The object is visually a single star and spectroscopically a double-line binary
candidate (Conti \& Eb\-bets 1977\nocite{1977ApJ...213..438C}; How\-arth et al.
1997\nocite{1997MNRAS.284..265H}). There is an IRAS source (\object{IRAS\,02245+5217})
located at 2 arcmin ($\rm \sim 2\,pc$) south of HD\,15137, that could be part
of a larger loop-like structure visible on the IRAS $60\,\mu$m image. No young
stellar clusters or stars earlier than B5 are reported within a 65\,pc radius from
HD\,15137.

{\it Our result:} The stellar density maps does not show evidence for a cluster
near HD\,15137. The two density peaks at coordinate offset (0,-2) in the TNG
density map (second row in Fig.\,\ref{f_hd3687}) are associated with an IR
source and could therefore be a physical cluster. This cluster is however not
detected on the 2MASS map.

%% file: STARS_TEX/hd36879.tex
\subsection{HD\,36879} 

Spectro\-scopically, the star is a single object that is found to have
un\-expectedly narrow Si\,{\sc iv} lines in its IUE spectra by Walborn \& Panek
(1984)\nocite{1984ApJ...286..718W}. These stellar wind lines change significantly
in the course of 4 days.  The presence of narrow Si\,{\sc iv} and
C\,{\sc iv} absorp\-tion are preferentially seen towards O-type stars in cluster and/or
nebulosities (de Kool \& de Jong 1985)\nocite{1985A&A...149..151D}.  These lines
are thought to be due to slowly eva\-porating clouds within the stellar wind
bubble. Interestingly, an isolated T\,Tauri star is reported to be located at a distance of
$\sim 10$\,pc  by Li \& Hu (1998\nocite{1998A&AS..132..173L}). The nearest early type star is \object{HD\,245310}, a B2
object, at a projected distance of $\rm \sim7\,pc$. No clusters are found within
65\,pc radius from the target star.

{\it Our result:} The NTT $\rm K_{s}$-band image shows that HD\,36879
consists of two objects with a separation of $9''.6$ and $\rm
\Delta K = 1.2^{\it m}$. HD\,36879 is therefore either a visual or an
optical binary. Cross-check with 2MASS finds that the companion
object has NIR colours $\rm (J-K) = 1.08$, indicating a late-type
object. The density maps indicate no physical cluster near this star. 


%% file: FIG_TEX/Fig4AP.tex
\begin{figure*}[ht]
 \center
\includegraphics[height=15cm,width=16cm,angle=-90]{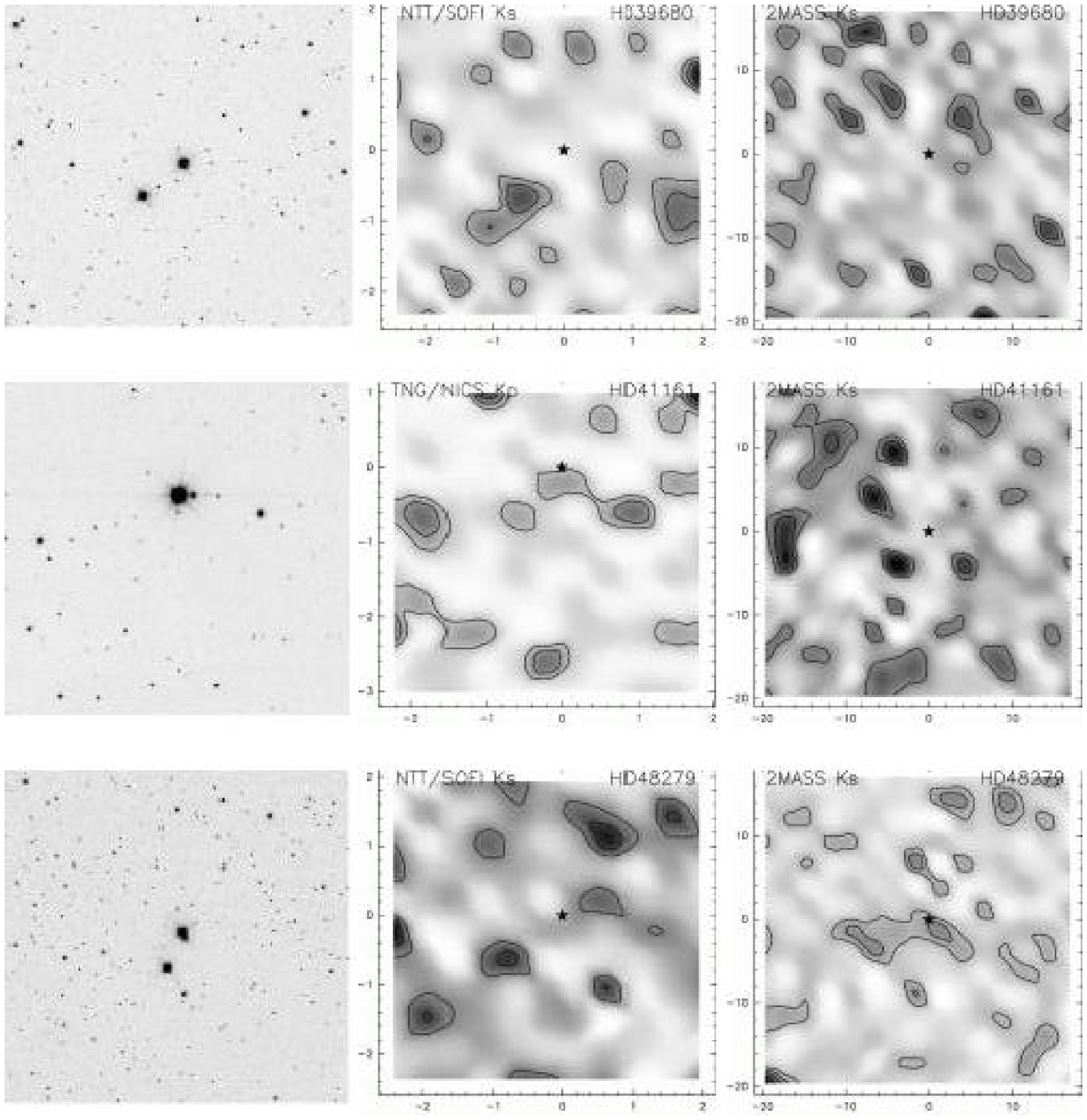}
\caption[]{As in Fig.\,\ref{f_hd3687} for stars HD\,39680, HD\,41161, and HD\,48279.}
\label{f_hd4827}
\end{figure*}

%% file: STARS_TEX/hd39680.tex
\subsection{HD\,39680}

HD\,39680 is an emission line star with double peaked Bal\-mer line emission (Gies
\& Bolton 1986\nocite{1986ApJS...61..419G}) and an infra\-red excess which is
interpreted as free-free emission originating in the stellar wind.  Mar\-chenko
et al. (1998\nocite{1998A&A...331.1022M})  notes the formidable long-term
photometric variation of this object resembling classical Be star behaviour. In
this sense it is similar to another Oe star in our sample: HD\,60848. HD\,39680 is
an optical binary (Lindroos 1985\nocite{1985A&AS...60..183L}) whose
comp\-onents were observed by M98 and appeared as single stars.  There are no
clusters within 65\,pc of HD\,39680 and the star is in fact the only O-type star
within a radius of 100 arcmin ($\rm \sim 70\,pc$). The closest early type star
is at a projected distance of 11\,pc, viz. the variable B4\,V star \object{Z\,Ori}. 

{\it Our result:} There is no indication of a physical cluster near this
 object from the density maps in Fig.\,\ref{f_hd4827}.


%% file: STARS_TEX/hd41161.tex
\subsection{HD\,41161}

The object forms a visual binary system with the component separated at
9.8\,arcsec (M98). Unexpectedly, Noriega-Crespo et al.
(1997\nocite{1997AJ....113..780N}) report a resolved IR bow shock near this
system that would be indicative of large spatial velocities. This visual binary is
relatively far from the Galactic plane.

{\it Our result:} The stellar densities near HD\,41161 are low. In the K-band image 
of the second row of Fig.\,\ref{f_hd4827}, the secondary star is detected to the
right of HD\,41161. The density maps do not reveal any physical cluster near
HD\,41161. 


%% file: STARS_TEX/hd48279.tex
\subsection{HD\,48279}

The multiple nature of HD\,48279 is visible in the left panel of the third row in
Fig.\,\ref{f_hd4827}. This system is likely to be an optical configuration (M98).
The small group of 4 objects is associated with an IR source:  \object{IRAS\,06400+0146}.
The target star lies in the field of the \object{Mon\,OB2} association, which has a
comparable but smaller distance of 1.6\,kpc (Mel'Nik \& Efremov
1995\nocite{1995AstL...21...10M}); HD\,48279's membership is not clear. The young
cluster \object{Dolidze 25} is found within 65\,pc of the target star. This
cluster however is located at a distance of 6.3\,kpc, much further away than HD\,48279.


{\it Our result:} The NTT $\rm K_{s}$-band density map shows no indication of
clustering of sources within a projected field of view of 1.4\,pc. The 2MASS
den\-sity map however shows at off\-set (-20,-20) a hint of a stellar
cluster. A visual inspection of a DSS image reveals a clustering of bright
objects at the same location. It appears to be centred on the A\,2 star 
\object{HD\,289038}.



%% file: FIG_TEX/Fig5AP.tex
\begin{figure*}[th]
 \center
 \includegraphics[height=15cm,width=16cm,angle=-90]{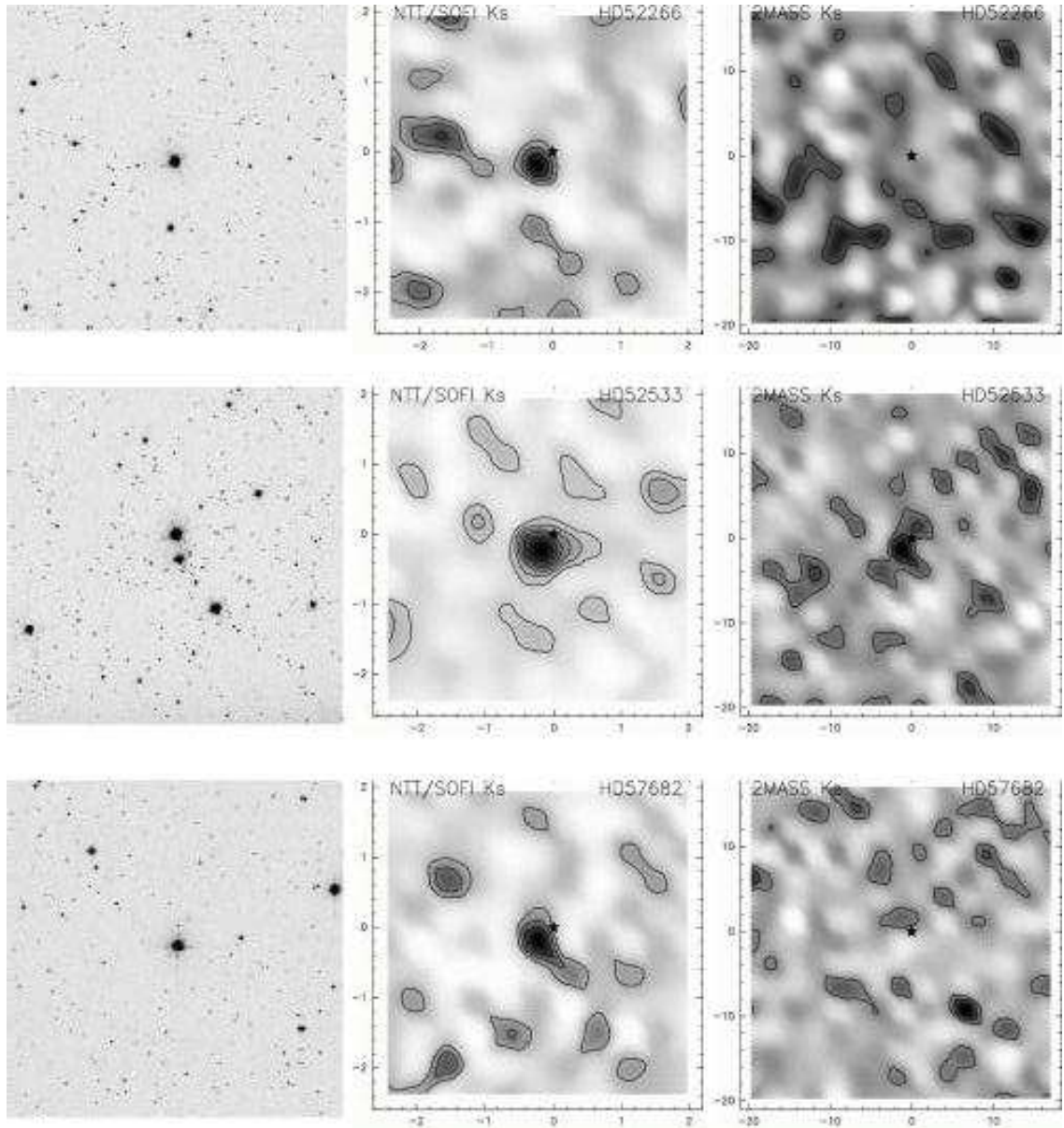}
\caption[]{As in Fig.\,\ref{f_hd3687} for stars HD\,52266, HD\,52533, and HD\,57682.}
\label{f_hd5768}
\end{figure*}

%% file: STARS_TEX/hd52266.tex
\subsection{HD\,52266}

The object is a candidate spectroscopic binary. It is projected close to the \object{CMa OB\,1} 
association that has a distance of 1\,kpc (Kalt\-cheva \& Hilditch
2000\nocite{2000MNRAS.312..753K}). HD\,52266's distance of 1.7\,kpc locates it further away. The Hip\-parcos distance estimate would put HD\,52266 at
485\,pc with considerable uncertainty. 

{\it Our result:} The high resolution stellar density map in the
middle panel of the upper row of Fig.\,\ref{f_hd5768} shows
evidence for a cluster near the target star. This cluster is not detected on the
2MASS density map.

%% file: STARS_TEX/hd52533.tex
\subsection{HD\,52533}
The visual multiplicity of the star is clearly visible in the
first panel of the second row of Fig.\,\ref{f_hd5768}. Part of
this multiple system (M98) is the early-type star \object{SAO\,134062} and
at 50 arcsec the B1 star HD\,52504). HD\,52533 itself is a single-line spectroscopic
binary with a period of 3.3 days (Gies \& Bolton 1986). It is
associated with a $60\,\mu$m excess (Noriega-Crespo et al. 1997)
and an unpublished X-ray source (\object{1RXS J070126.3-030704}).

{\it Our result:} Both the density map derived from the $\rm K_{s}$-band NTT image
and the 2MASS map in Fig.\,\ref{f_hd5768} show a strong indication for clustering of stars
near HD\,52533.


%% file: STARS_TEX/hd57682.tex
\subsection{HD\,57682} 

HD\,57682 is spectroscopically a single object. It is associated with the
unpublished X-ray source \object{1RXS J072201.7-085844} and with the IR source \object{IRAS
07196-0852}. The star is reported to be a variable Oe star by Cot\'{e} \& van
Kerkwijk 1993\nocite{1993A&A...274..870C} (see also MacConnell
1981)\nocite{1981A&AS...44..387M}. Having a rather small $v{\rm sin}i$ of 33\,$\rm
km\,s^{-1}$ (Penny 1996)\nocite{1996ApJ...463..737P}, the object's rotational axis
is oriented probably at a near pole-on angle. The object is suggested to be a
runaway star by Comeron et al. (1998)\nocite{1998A&A...330..975C}.


{\it Our result:} The NTT stellar density map shows signs of a
cluster near HD\,57682. The 2MASS density map however only gives evidence for a    
possible cluster centred at offset (6,-9). SIMBAD and DSS do not
provide additional indications for a cluster at this position.  


%% file: FIG_TEX/Fig6AP.tex
\begin{figure*}[ht]
 \center
 \includegraphics[height=15cm,width=16cm,angle=-90]{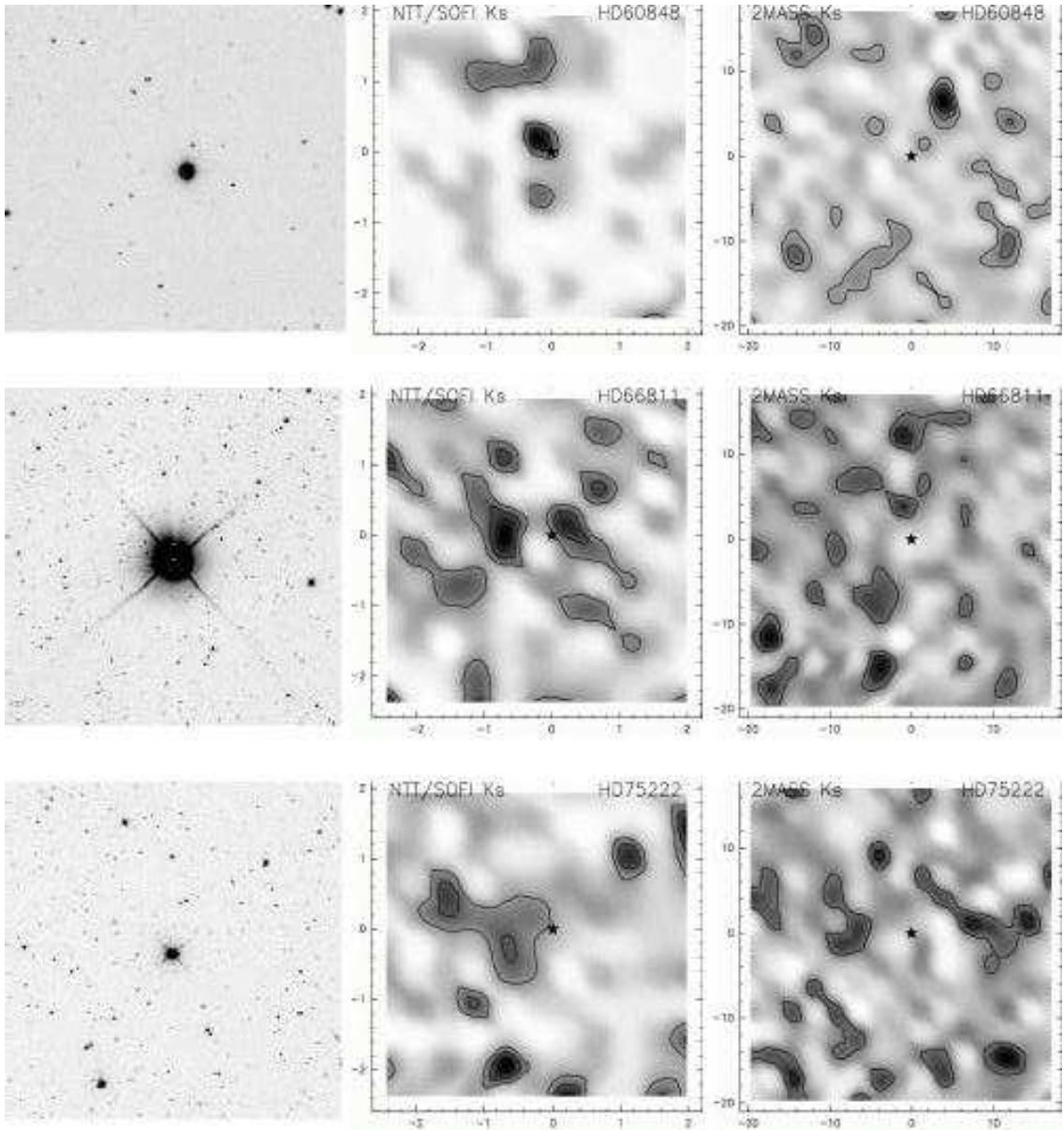}
\caption[]{As in Fig.\,\ref{f_hd3687} for stars HD\,60848, HD\,66811, and HD\,75222.}
\label{f_hd7522}
\end{figure*}

%% file: STARS_TEX/hd60848.tex
\subsection{HD\,60848 ($\equiv$BN\,GEM)} 

BN Gem is a star of unknown spectroscopic multiplicity status. The object has an
emission line spectrum and manifests similar continuum emission variability as classical
Be stars (Divan, Zorec \& Andrillat 1983\nocite{1983A&A...126L...8D}, see also HD\,39680) 
The object has a reported $v {\rm sin}i = 240\,{\rm km\,s^{-1}}$ (Penny 1996).
There are no clusters, O-type or early B-type stars within a projected radius of
65\,pc from HD\,60848.

{\it Our result:} The stellar density map derived from the NTT image does not show
stellar density peaks near this object. The 2MASS density map shows a density peak
at offset (4,5) that is likely to be statistical. 
 

%% file: STARS_TEX/hd66811.tex
\subsection{HD\,66811 ($\equiv \zeta$ Pup)}

$\rm \zeta$ Pup is the nearest O-type star to the sun and a runaway object (e.g. 
Upton 1973\nocite{1973gnrp.conf..119U}), although its radial velocity is
insufficient to be refer to by M98 and G87 as such (see Sect.\,\ref{sample}).  It
is spectroscopically single and one of the exciting stars of the \object{Gum Nebula} (Chanot
\& Sivan 1983\nocite{1983A&A...121...19C}).  Identification of $\zeta$ Pup with a
parent cluster/OB-association has failed so far (see Hooger\-werf et al. 2001,
Van\-beveren et al. 1998\nocite{1998brbi.book.....V}). The star is included for
completeness reasons as it is part of the M98 sample of field O stars.

{\it Our result:} The density maps show the expected absence of a cluster centred on 
the star. The NTT/SOFI image has a density peak at an offset of (0,-0.6).

%% file: STARS_TEX/hd75222.tex
\subsection{HD\,75222}

Hipparcos measured this star being a runaway object with a peculiar space velocity
of $\rm 57.2\,km\,s^{-1}$ (Hoogerwerf et al. 2001). It is listed as a candidate
member of the \object{Vel OB1} association by Reed (2000\nocite{2000AJ....119.1855R}) on the
basis of an extinction study. Like $\rm \zeta$~Pup, the star is included in our
study for completeness reasons (see Sect.\,\ref{sample}). 


{\it Our result:} The density maps show the expected absence of clusters.

%% file: FIG_TEX/Fig7AP.tex
\begin{figure*}[ht]
 \center
 \includegraphics[height=15cm,width=16cm,angle=-90]{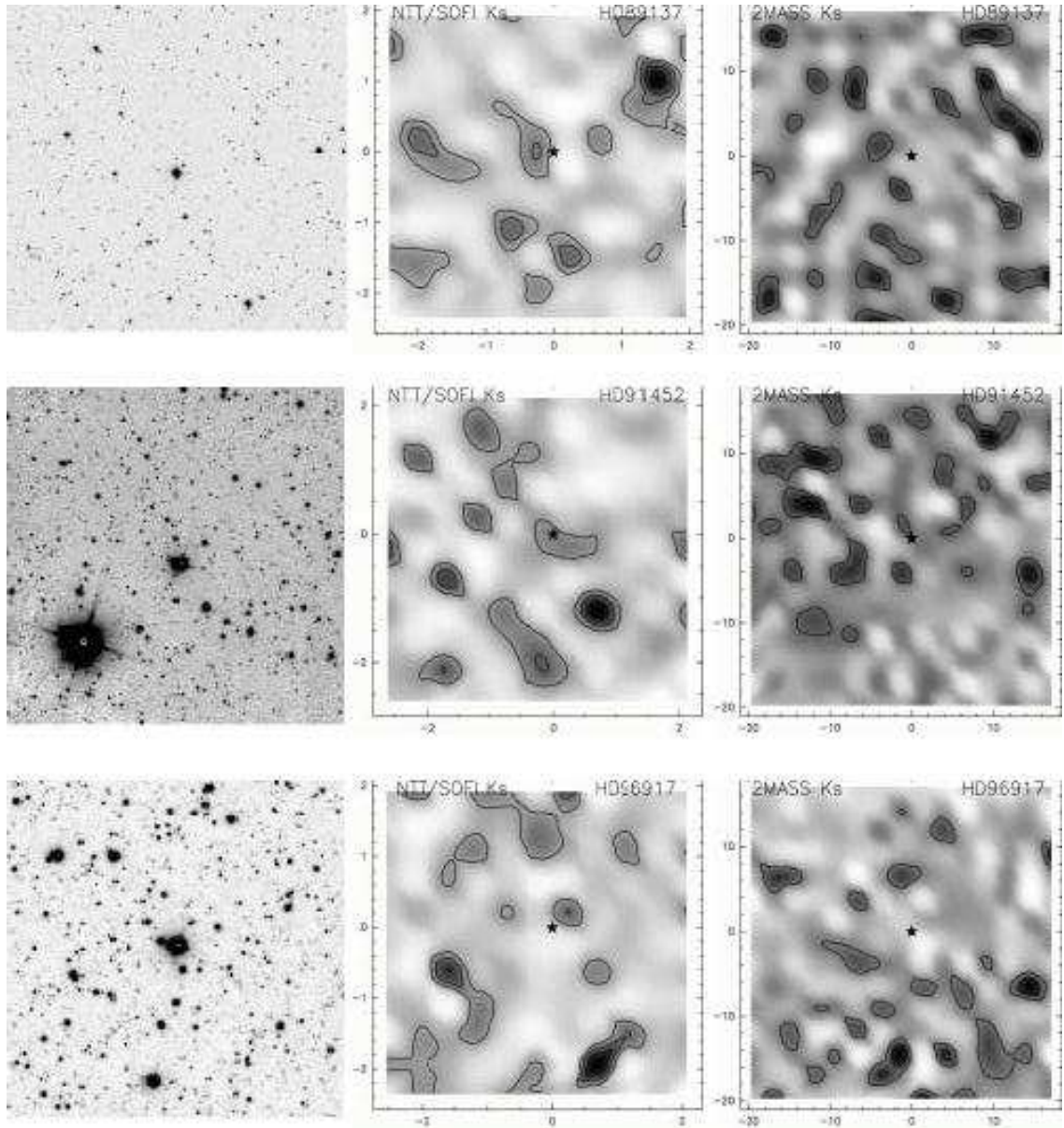}
\caption[]{As in Fig.\,\ref{f_hd3687} for stars HD\,89137, HD\,91452, and HD\,96917.}
\label{f_hd9691}
\end{figure*}

%% file: STARS_TEX/hd89137.tex
\subsection{HD\,89137}

The star is reported to have a peculiar spectrum by Walborn
(1976\nocite{1976ApJ...205..419W}) and was found to be a candidate single-line
spectroscopic binary by Levato et al. (1988)\nocite{1988ApJS...68..319L}. The
source \object{IRAS\,10137-5059} is located at 1.3 arcmin. This object is unresolved at
the 1 arcmin resolution IRAS maps of Noriega-Crespo et al. 
(1997)\nocite{1997AJ....113..780N}. Extended $60\,\mu$m~emission is detected by
IRAS and comprises also HD\,89137. 


{\it Our result:} The stellar density map derived from the NTT image in
Fig.\,\ref{f_hd9691} shows a strong density peak at offset (1.6,1.0). This cluster
is not associated with any known physical object in SIMBAD.


%% file: STARS_TEX/hd91452.tex
\subsection{HD\,91452}

The object is spectroscopically a single star, that is located in the
direction of the Carina Spiral Feature and at a 1.3 degrees distance
from the Theta Carina cluster. HD\,91452 suffers from a reasonable
extinction of $\rm A_{V}\simeq 1.5^{\it m}$.  There are no known
clusters within a projected radius of 65\,pc of HD\,91452.  The
nearest early type star is a B5 star HD\,307781 at 37\,pc.


{\it Our result:} The stellar density maps do not show any cluster near HD\,91452.

%% file: STARS_TEX/hd96917.tex
\subsection{HD\,96917}

The object is a candidate single-line spectroscopic binary and a known variable
star with evidence for photometric mic\-ro\--var\-iabi\-lity (Balona
1992\nocite{1992MNRAS.254..404B}).  Like HD\,91452, HD\,96917 is located in the direction of the Carina Spiral Feature. The
nearest early-type star is the B2 star \object{HD\,96088} at a distance of 51\,pc.

{\it Our result:} The stellar density maps do not reveal any physical cluster.



%% file: FIG_TEX/Fig8AP.tex
\begin{figure*}[ht]
 \center
 \includegraphics[height=15cm,width=16cm,angle=-90]{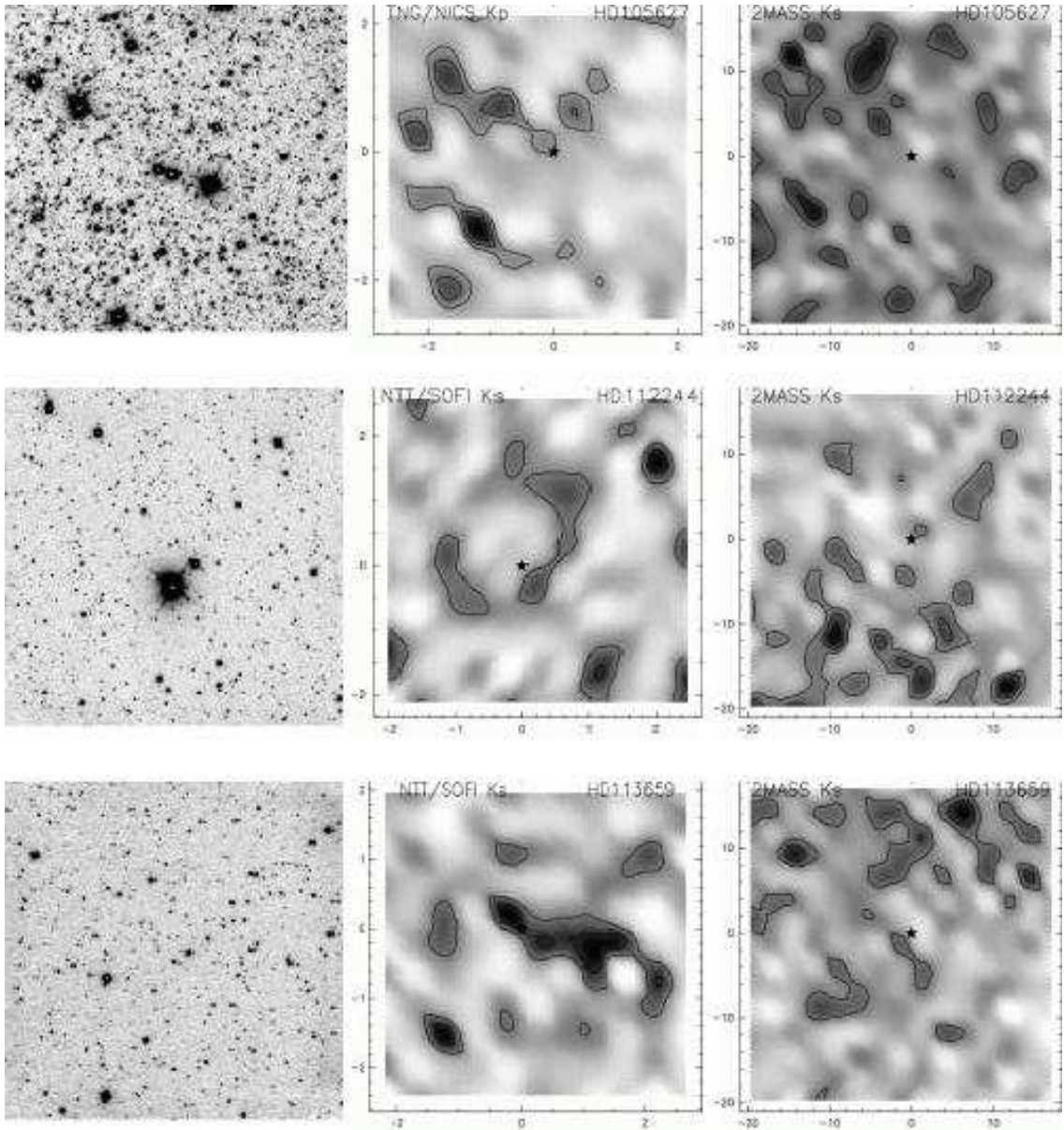}
\caption[]{As in Fig.\,\ref{f_hd3687} for stars HD\,105627, HD\,112244, and HD\,113659.}
\label{f_hd1136}
\end{figure*}

%% file: STARS_TEX/hd105627.tex
\subsection{HD\,105627} 

HD\,105627 is located in a crowded field, but isolated from other O-type stars.
The nearest O-type star is in fact at an angular distance of 1.44 degrees (80\,pc). It
is a relatively little studied star, with an Hipparcos astrometric optical
companion at 14\,arcsec (M98).  


{\it Our result:} The density maps do not indicate any physical
cluster.

%


%% file: STARS_TEX/hd112244.tex
\subsection{HD\,112244}

This emission line object is a visual binary from the Lindroos
sample (Lindroos 1985\nocite{}). The secondary is a K0III star.
Recently, the system was suggested to be an optical
configuration (Hu\'{e}lamo et al. 2000)\nocite{2000A&A...359..227H}). The primary O-type star is a
candidate single-line spectroscopic binary. HD\,112244 shows
regular photometric variability with three possible 
periods (Marchenko et al. 1998). It is associated with the IR
source \object{IRAS\,12529-5633} and it is an X-ray source. The closest
early type-star is another emission line O star \object{HD\,112147}, at
a projected distance of 57\,pc.


{\it Our result:} The density maps in Fig.\,\ref{f_hd1136} do not reveal any physical
clustering near HD\,112244. The 2MASS map density peaks are likely to be statistical.


%% file: STARS_TEX/hd113659.tex
\subsection{HD\,113659}

A star with unknown spectroscopic multiplicity status according to
M98, although it is reported to have a variable radial velocity
and to be a member of the Cen\,OB1 association according to the
LSC (Humphreys \& McElroy 1984\nocite{1984ApJ...284..565H}; see
also Mathys 1988\nocite{1988A&AS...76..427M}). Therefore this star is a doubtful
field O star.


{\it Our result:} The stellar density map do not show clusters.


%% file: FIG_TEX/Fig9AP.tex
\begin{figure*}[ht]
 \center
 \includegraphics[height=15cm,width=16cm,angle=-90]{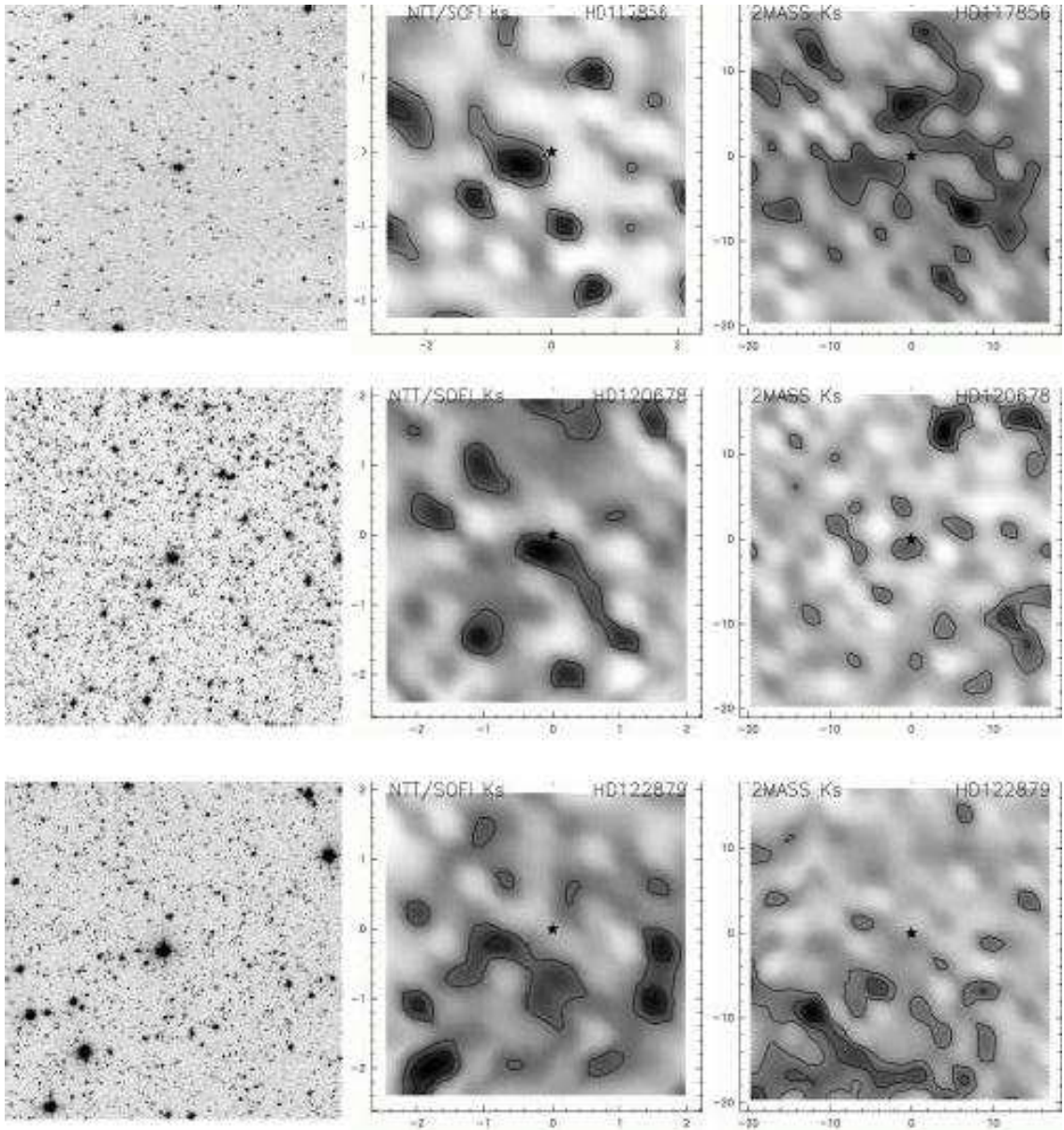}
\caption[]{As in Fig.\,\ref{f_hd3687} for stars HD\,117856, HD\,120678, and HD\,122879.}
\label{f_hd1228}
\end{figure*}

%% file: STARS_TEX/hd117856.tex
\subsection{HD\,117856}
The object is a close visual binary (1.6\,arcsec) and a candidate
double-line spectroscopic binary. 
The field shows a collection of interstellar dark clouds and H\,{\sc ii}
regions, that may be part of the \object{Southern Coalsack} (distance of
180\,pc). With an estimated distance of 1.7\,kpc, HD\,117856 is
probably piercing through the fringes of this star formation region. 
\object{Stock\,16} is a young cluster $\rm \sim60\,pc$ away and related to the H\,{\sc ii}
region \object{RCW\,75}. It has a distance of $\rm 1.9\,kpc$ (Turner
1985\nocite{1985ApJ...292..148T}), comparable to that of HD\,117856.

{\it Our result:} The density maps do not reveal any physical clusters.

\begin{table}[h]
 {
 \begin{center}
  \begin{tabular*}{0.49\textwidth}
   {@{\extracolsep{\fill}}lccccc}
\hline
\hline
 Cluster Name & Ang Dist & Lin Dist & Age & Dist & Ref \\
	& (deg)	& (pc) & (Myr) & (kpc) & \\
\hline
Stock 16                  & 1.9 &57 & 4 & $1.9\pm0.08$ & 1\\
 \hline
 \end{tabular*}
 \end{center}
\vspace{-0.2cm}
{\tiny 1: Turner 1985\nocite{1985ApJ...292..148T} }
}
\end{table}


%% file: STARS_TEX/hd120678.tex
\subsection{HD\,120678} 

The object is a variable emission line star and a rapid rotator with $v{\rm sin}i =
350\,{\rm kms^{-1}}$ (Conti \& Ebbets 1977\nocite{1977ApJ...213..438C}). HD\,120678
is of unknown spectroscopic multiplicity status. It is located in a crowded field
together with two other early-type stars: a B2 star at 4.5\, pc (\object{HD\,120634}), and a
B5 star at 3.3\,pc (\object{HD\,120578}). 

{\it Our result:} The stellar density maps do not show any
physical cluster.

%% file: STARS_TEX/hd122879.tex
\subsection{HD\,122879}
Spectroscopically a single star that has been noted as a variable object with a
period of 1.58 days based on Hipparcos photometry by Marchenko et
al. (1998)\nocite{1998A&A...331.1022M}.  The spectral type is probably B0Ia rather than 
O9.5I as listed in Table\,1 (following Garrison et
al. 1977\nocite{1977ApJS...35..111G}; Walborn \& Fitzpatrick
1990)\nocite{1990PASP..102..379W}. The object was suggested to be member of 
\object{Cen OB1} (e.g. Pawlowicz \& Herbst 1980)\nocite{1980A&A....86...68P}.


{\it Our result:} The stellar density maps do not show any
clustering near this object. 



%% file: FIG_TEX/Fig10AP.tex
\begin{figure*}[ht]
 \center
 \includegraphics[height=15cm,width=16cm,angle=-90]{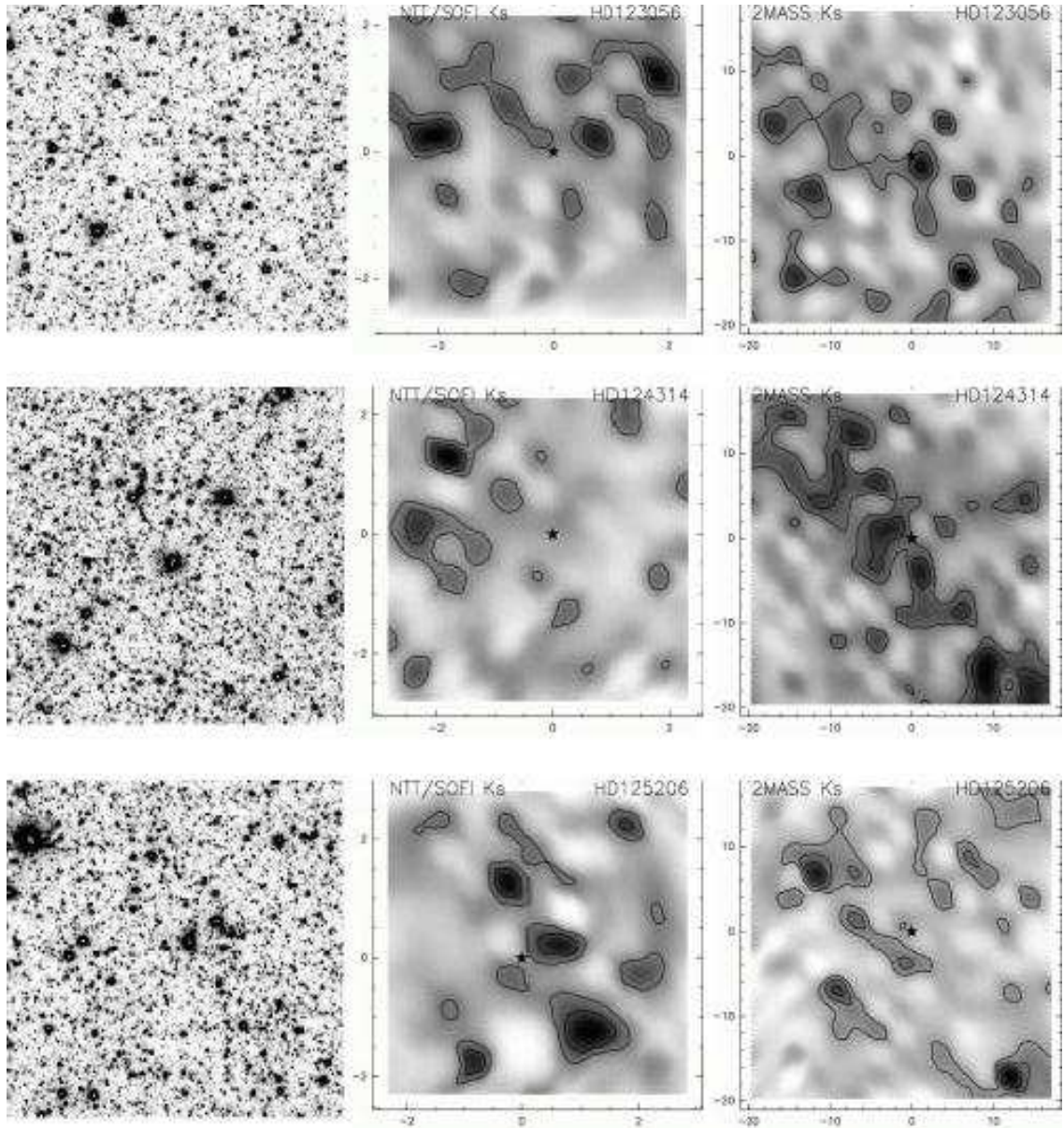}
\caption[]{As in Fig.\,\ref{f_hd3687} for stars HD\,123056, HD\,124314, and HD\,125206.}
\label{f_hd1252}
\end{figure*}

%% file: STARS_TEX/hd123056.tex
\subsection{HD\,123056} The object is spectroscopically a single star and not part
of the G87 sample of O-type field stars. Little is known about it. An IR
source (\object{IRAS\,14040-6014}) is located at 1.5\,arcmin SE of HD\,123056. 



{\it Our result:} The stellar density maps do not show a physical cluster near
this object. 


%% file: STARS_TEX/hd124314.tex
\subsection{HD\,124314}
The object is a visual binary star with a 2.7\,arcsec separation.
It is a candidate single-line spectroscopic binary (Feast et al.
1955\nocite{1955MmRAS..67...51F}), although Balona
(1992\nocite{1992MNRAS.254..404B}) finds a nearly
constant light curve. HD\,124314 is an emission line star and the
main ionizing source of the H\,{\sc ii} region \object{RCW\,85} (Yamaguchi
et al. 1999). It is associated with the ROSAT source
\object{1RXS\,J141500.1-614231} and the IR source (1.5 arcmin NW)
\object{IRAS\,14111-6127}. The young cluster \object{NGC\,5606} is at a projected linear 
distance of $\rm \sim 45\,pc$, but roughly twice as far from the sun than
HD\,124314.


{\it Our result:} The density maps do not reveal any stellar cluster associated with this star.

%% file: STARS_TEX/hd125206.tex
\subsection{HD\,125206} 

A candidate double-line spectroscopic binary. The star is located $\rm
\sim10\,pc$ north of the star forming H\,{\sc ii} region \object{RCW\,85} (see
Yamaguchi et al.  1999)\nocite{1999PASJ...51..765Y}. This region is at a similar
distance as HD\,125206.  Mel'Nik \& Efremov (1995) note that HD\,125206 may
belong to their {\it Clust\,3} group. HD\,125206 is at a distance 
comparable to the one of the young cluster \object{NGC\,5606}. The projected
distance between these two objects is less than $\rm 65\,pc$.


{\it Our result:} The stellar density maps do not show an indication of a cluster
near HD\,125206. The 2MASS density peak at offset (-12,7) in Fig.\,\ref{f_hd1252}
could be associated with the IR source \object{IRAS\,14183-6044}. Another stellar
density peak at offset (12,-17) lies close to the IR source 
\object{IRAS\,14143-6108}. 


\begin{table}[h]
 {
 \begin{center}
  \begin{tabular*}{0.49\textwidth}
  {@{\extracolsep{\fill}}lccccc}
\hline
\hline
 Cluster Name & Ang Dist & Lin Dist & Age & Dist &Ref\\
	& (deg) & (pc) & (Myr) & (kpc) & \\
\hline
  NGC5606    & 1.7  &51    &  7  & $2.4\pm0.2$ &  1\\
\hline
 \end{tabular*}
 \end{center}
 \vspace{-0.2cm}
{\tiny 1: Vazquez et al. (1994)\nocite{1994A&AS..106..339V}}
 }
\end{table}



%% file: FIG_TEX/Fig11AP.tex
\begin{figure*}[ht]
 \center
 \includegraphics[height=15cm,width=16cm,angle=-90]{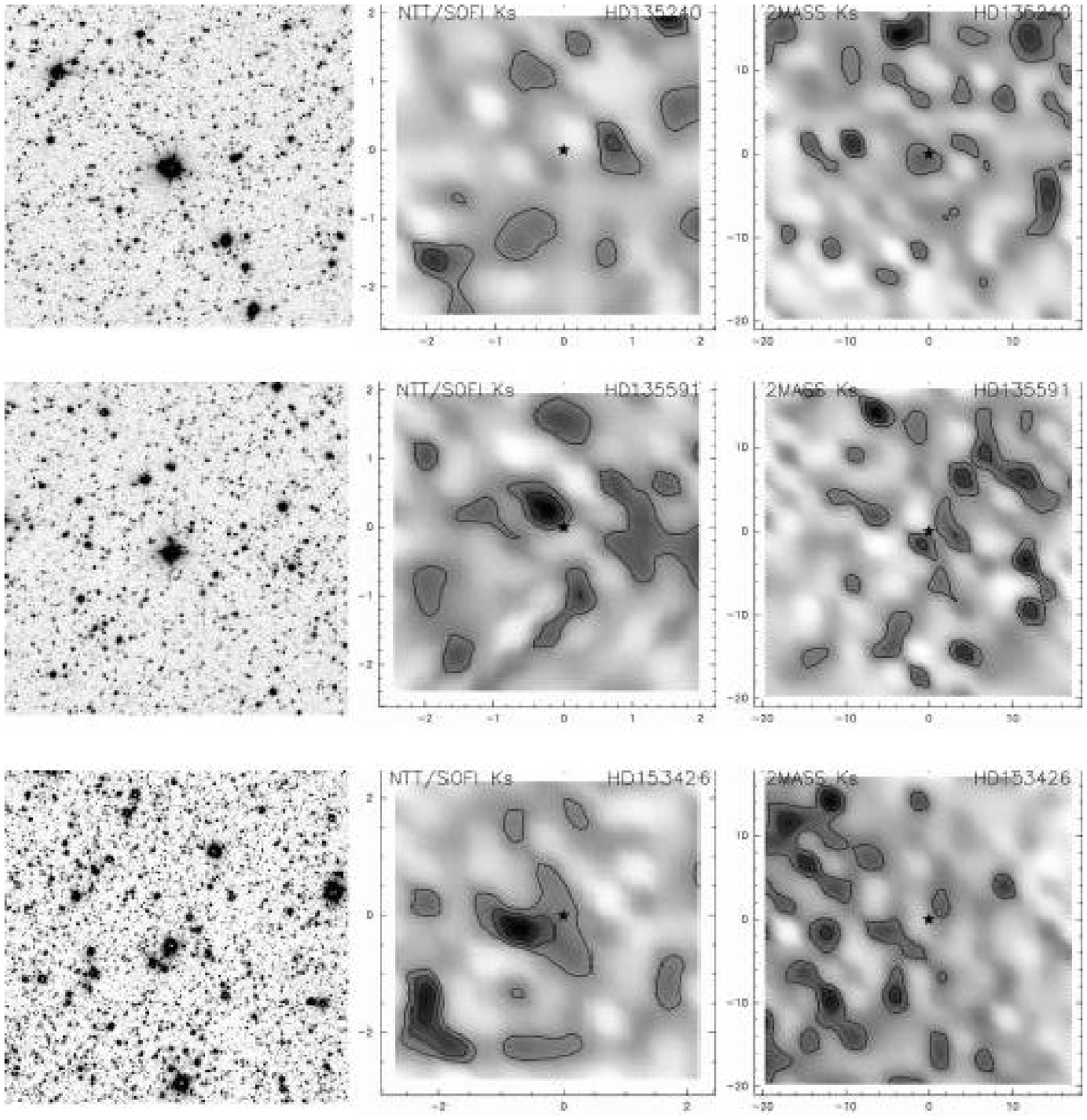}
\caption[]{As in Fig.\,\ref{f_hd3687} for stars HD\,135240, HD\,135591, and HD\,153426.}
\label{f_hd1534}
\end{figure*}

%% file: STARS_TEX/hd135240.tex
\subsection{HD\,135240}
This double-line ellipsoidal spectroscopic binary is shown to be
a massive triple system consisting of  O7III-V, O9.5V, and 
B0.5V stars (Penny et al. 2001)\nocite{2001ApJ...548..889P}. The
primary and secondary have different evolutionary ages of 2.5\,Myr
and 5.1\,Myr, respectively. The source is associated with the X-ray
source 1RXS J151658.5-605730. Another early-type Be star
(HD\,135160) is located at 1\,pc. A third early-type B3 star is at
a distance of 1.3\,pc. HD\,135240 may belong to Mel'Nik \&
Efrimov's (1995) {\it Pis\,20} group.

{\it Our result:} The density maps do not indicate any cluster near this triple
system.


%% file: STARS_TEX/hd135591.tex
\subsection{HD\,135591}
A triple object of which the third component is located at 44.5
arcsec (Lindroos 1985).  The latter is an A8\,III star, that would
still be in the pre-main sequence phase. HD\,135591 is also
associated with the X-ray source \object{1RXS\,J151848.4-602952}. This
system may belong to Mel'Nik \& Efrimov's (1995) {\it Pis\,20}
group. There are a number of nearby early-type stars, e.g.
\object{HD\,135786} at $\rm \sim3\,pc$, but known young clusters are
absent within 65\,pc radius.


{\it Our result:} There is no indication of a cluster near this star.

%% file: STARS_TEX/hd153426.tex
\subsection{HD\,153426}

The object is a candidate double-line spectroscopic binary and located within a half
degree of the H\,{\sc ii} region Sharpless\,2. It was suggested by Walborn
(1973)\nocite{1973AJ.....78.1067W} that the object could be physically related to
the runaway HMXB HD\,153919/4U1700-37 (see Ankay et al.
2001\nocite{2001A&A...370..170A}), that is at a projected linear distance of $\rm
\sim24\,pc$.  Note that SIMBAD lists HD\,153426 wrongly as a B9II-III star. 

{\it Our result:} The density map in Fig.\,\ref{f_hd1534} derived
from NTT imaging shows the target star associated with a clear
density peak. Although HD\,153426 is offset from the cluster
centre, we consider the density peak as a physical cluster.



%% file: FIG_TEX/Fig12AP.tex
\begin{figure*}[ht]
 \center
 \includegraphics[height=15cm,width=16cm,angle=-90]{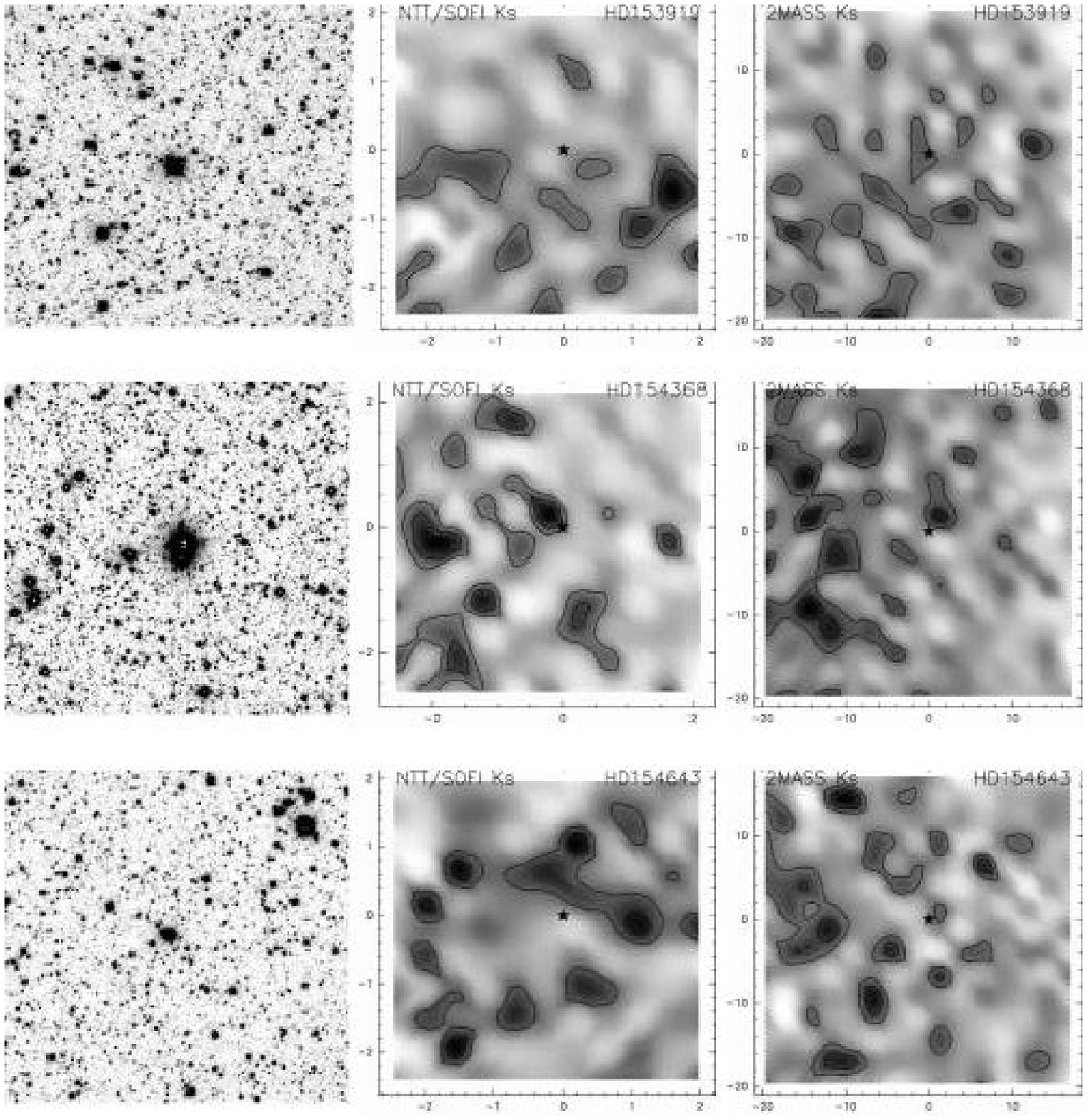}
\caption[]{As in Fig.\,\ref{f_hd3687} for stars HD\,153919, HD\,154368, and HD\,154643.}
\label{f_hd1546}
\end{figure*}

%% file: STARS_TEX/hd153919.tex
\subsection{HD\,153919}
The object is a single-line eclipsing spectroscopic binary and
known as the runaway HMXB 4U1700-37 (Ankay et al. 2001). The
object is located in the field of the open cluster \object{NGC\,6281}, 
also known as NGC6281-2. It is included in this survey
for completeness reasons as it is part of the M98 subsample of O-type
field stars (see Sect.\,\ref{sample}).

{\it Our result:} The density maps show the expected absence of clusters near the object.

%% file: STARS_TEX/hd154368.tex
\subsection{HD\,154368}

HD\,154368 is a visual binary and an eclipsing binary with a 16.1\,day period,
located near the \object{Sco OB\,1} association.  The spectroscopic distance
estimate is 800\,pc (Snow et al. 1996\nocite{1996ApJ...465..245S}).
The Hipparcos parallax measurement allows an uncertain distance estimate of 370\,pc.
HD\,154368 is mainly known for a translucent cloud in its direction. Its
coordinates puts it within the error ellipse of the IR source \object{IRAS\,17031-3522}. The
young cluster \object{Bochum\,13} is located within 65\,pc projected radius. According to
Battinelli et al. (1994) the error on the absolute distance modulus can 
be $\rm \sim 0.5^{m}$. Therefore the distance to the Sun of HD\,154368 and
Bochum\,13 are similar within the adopted errors.

{\it Our result:} The target star is not found in a cluster on the density maps in
Fig.\,\ref{f_hd1546}.


\begin{table}[h]
 {
 \begin{center}
  \begin{tabular*}{0.49\textwidth}
   {@{\extracolsep{\fill}}lccccc}
\hline
\hline
   Cluster Name & Ang Dist & Lin Dist & Age & Dist & Ref \\
	& (deg) & (pc) & (Myr) & (kpc) & \\
\hline
Bochum\,13 &  2.2 &42  & 6.3 & $1.7\pm0.45$  & 1\\
\hline
 \end{tabular*}
 \end{center}
 \vspace{-0.2cm}
 {\tiny 1: Battinelli et al. (1994)\nocite{1994A&AS..104..379B}}
 }
\end{table}


%% file: STARS_TEX/hd154643.tex
\subsection{HD\,154643} 

HD\,154643 is a little studied, candidate single-line spectroscopic binary. Within
65\,pc, the young cluster \object{Bochum\,13} is found. It has an age of 6.3\,Myr and a
distance of 1.7\,kpc (Battinelli et al. 1994\nocite{1994A&AS..104..379B}). The
distance to the Sun of HD\,154643 is comparable to that of Bochum\,13.

{\it Our result:} The density maps do not show any indication of a physical cluster.


\begin{table}[h]
 {
 \begin{center}
  \begin{tabular*}{0.49\textwidth}
  {@{\extracolsep{\fill}}lccccc}
\hline
\hline
 Cluster Name & Ang Dist & Lin Dist & Age & Dist & Ref \\
	& (deg) & (pc) & (Myr) & (kpc) & \\
\hline
Bochum 13 & 1.9 & 44  & 6.3 & $1.7\pm0.45$  & 1\\
\hline
 \end{tabular*}
 \end{center}
 \vspace{-0.2cm}
 {\tiny 1: Battinelli et al. (1994)\nocite{1994A&AS..104..379B}}
 }
\end{table}

%% file: FIG_TEX/Fig13AP.tex
\begin{figure*}[ht]
 \center
 \includegraphics[height=15cm,width=16cm,angle=-90]{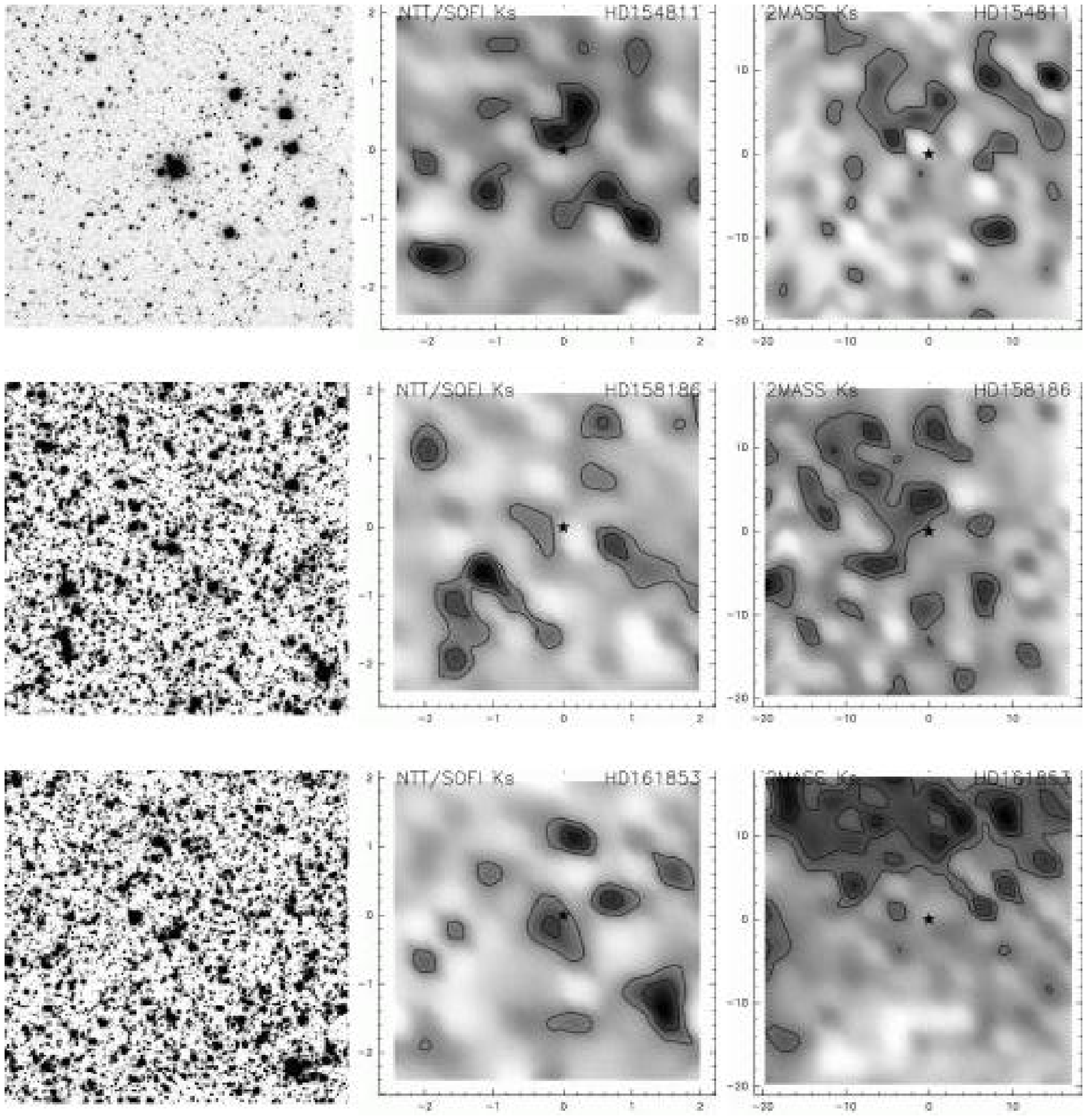}
\caption[]{As in Fig.\,\ref{f_hd3687} for stars HD\,154811, HD\,158186, and HD\,161853.}
\label{f_hd1618}
\end{figure*}

%% file: STARS_TEX/hd154811.tex
\subsection{HD\,154811}

HD\,154811 is most likely a single component supergiant (Levato et al. 1988).  An
uncertain Hipparcos distance would put it at 0.42\,kpc. An IR source
(\object{IRAS\,17060-4657}) is located at 1.5\,arcmin from HD\,154811, but the star falls
outside the IRAS error-ellipse.


{\it Our result:} Although the NTT $\rm K_{s}$-band image in Fig.\,\ref{f_hd1618} shows
some conspicuous bright stars, both density maps do not reveal any clustering near the
star. Whether this group of stars is associated to the field O star is unlikely, as
2MASS data shows that a number of them are bright K-band objects with $\rm J-K>1.5$, occuping 
the gaint branch region in a colour-magnitude diagram. This suggests that these are evolved 
low-mass stars.


%% file: STARS_TEX/hd158186.tex
\subsection{HD\,158186} 

This field star is a binary system whose eclipsing Hipparcos light curve was
presented by Mar\-chenko et al. (1998). The field itself shows signs of active star
formation.  HD\,158186 is associated with an unresolved IR source \object{IRAS\,17260-3129}
(Noriega-Crespo et al. 1997). The centre of the dark cloud LDN\,1732 coincides with
the position of the target star, which could also be the illuminating source of the
close by H\,{\sc II} region \object{BBW\,32300}.  On angular scales of 40 arcmin, clear
nebulosities and obscured regions are distinguishable from optical images, among
which the emission nebulae \object{Sh\,2-13} and \object{RCW\,133}. Three young clusters are found
within a projected radius of 65\,pc, among which \object{NGC\,6383} belonging to
\object{Sgr OB1} association with an estimated age of 1.7\,Myr. NGC\,6383 is at 
1.5\,kpc from the Sun (Fitzgerald et al. 1978)\nocite{}, comparable to HD\,158186's distance.

{\it Our result:} The density maps in Fig.\,\ref{f_hd1618} do not
show a cluster near the target star.

\begin{table}[h]
 {
 \begin{center}
  \begin{tabular*}{0.49\textwidth}
   {@{\extracolsep{\fill}}lccccc}
\hline
\hline
 Cluster Name & Ang Dist  & Lin Dist &  Age & Dist & Ref \\
	& (deg) & (pc) & (Myr) & (kpc) & \\
\hline
%
 NGC 6383         & 1.6 &30 & 1.7 & $1.5\pm0.2$ & 1\\
%
%
%
%
%
%
%
\hline
 \end{tabular*}
 \end{center}
 \vspace{-0.2cm}
 {\tiny 1: Fitzgerald et al. (1978)\nocite{1978MNRAS.182..607F}}; 
 }
\end{table}


%% file: STARS_TEX/hd161853.tex
\subsection{HD\,161853}

The target star is a candidate single-line spectroscopic binary and the exciting
star of the H\,{\sc ii} region \object{RCW\,134} (Yamaguchi et al. 1999\nocite{}). The
object has been listed as a candidate PN (Ratag et al.
1990\nocite{1990A&A...233..181R}) based on its IRAS colours and radio continuum
emission. HD\,161853 is also associated with the X-ray source
\object{1RXS\,J174916.5-311509}. Within 65\,pc radius there is the stellar cluster
\object{Collinder\,347} with an age less than 10\,Myr. Collinder\,347 has a
distance of 1.5\,kpc, comparable to that of HD\,161853.



{\it Our result:} The density map derived from the NTT image in
the bottom row of Fig.\,\ref{f_hd1618} does not reveal a cluster
near HD\,161853. The stellar density map derived from 2MASS shows
the effect of large scale extinction in the field. Although in a
star formation field, HD\,161853 does not seem to be associated
with a cluster.

\begin{table}[h]
 {
 \begin{center}
  \begin{tabular*}{0.49\textwidth}
   {@{\extracolsep{\fill}}lccccc}
\hline
\hline
 Cluster Name  & Ang Dist  & Lin Dist  & Age & Dist & Ref \\
	& (deg) & (pc) & (Myr) & \\
\hline
 Collinder 347 & 2.0 &55 & 6.3 & $1.5\pm0.35$ & 1\\
\hline
 \end{tabular*}
 \end{center}
\vspace{-0.2cm}
 {\tiny 1: Battinelli et al. (1994)\nocite{1994A&AS..104..379B}}
 }
\end{table}




%% file: FIG_TEX/Fig14AP.tex
\begin{figure*}
 \center
 \includegraphics[height=15cm,width=16cm,angle=-90]{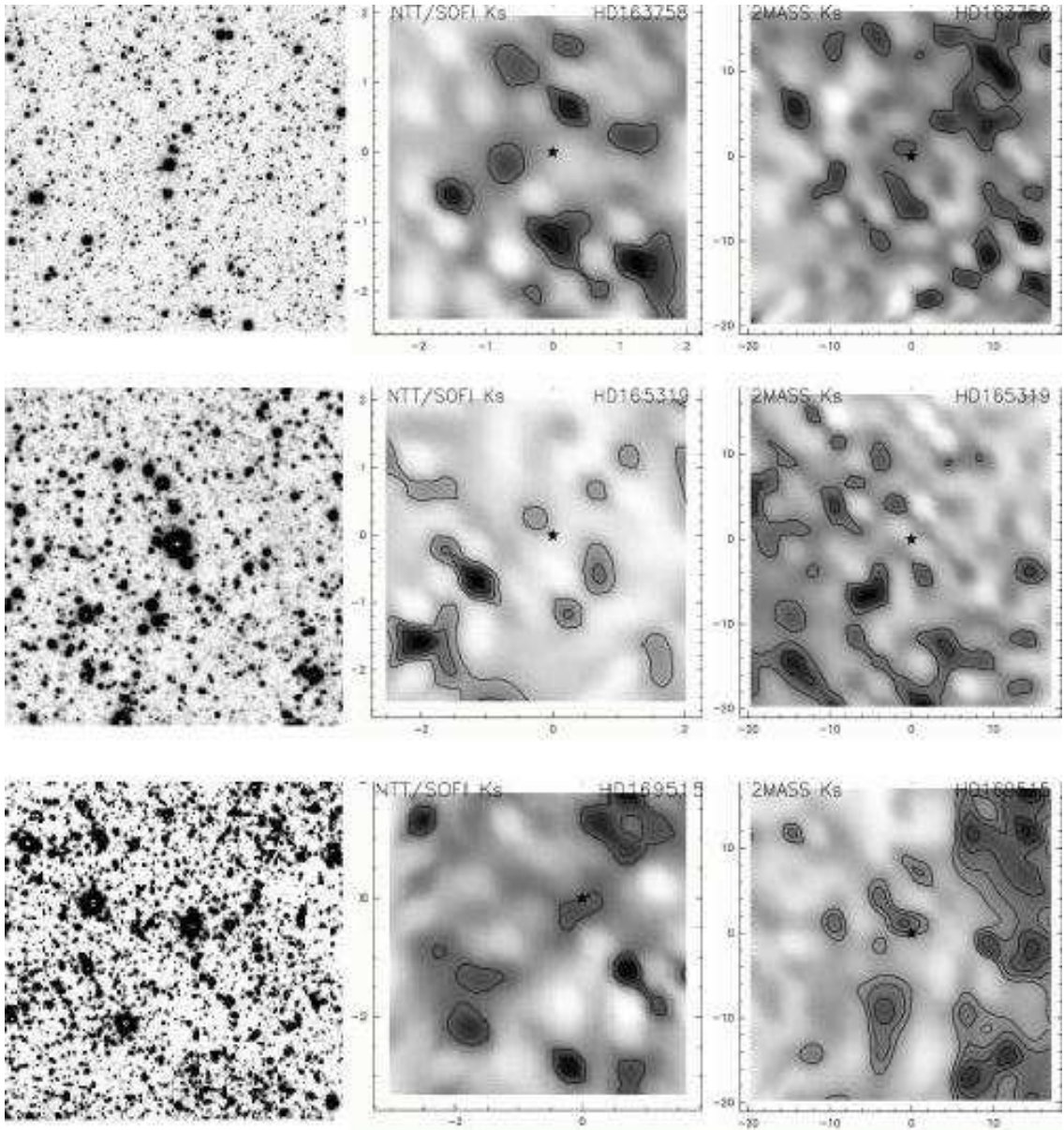}
\caption[]{As in Fig.\,\ref{f_hd3687} for stars HD\,163758, HD\,165319, and HD\,169515.}
\label{f_hd1695}
\end{figure*}

%% file: STARS_TEX/hd163758.tex
\subsection{HD\,163758}
The object is a single component Wolf-Rayet star located in an
average stellar field. The closest early-type star to HD\,163758
is the B\,2 object \object{HD\,163924} at $\sim 25$\,pc. No young clusters
are found within $\rm 65\,pc$ radius.

{\it Our result:} The density maps do not reveal any cluster near the target star.


%% file: STARS_TEX/hd165319.tex
\subsection{HD\,165319}

HD\,165319 is projected on the $\sim 23\times23$ arcmin H\,{\sc
ii} region \object{RCW\,158} (Rodgers et al.
1960\nocite{1960MNRAS.121..103R}). The object suffers from some
extinction, showing a  colour excess of $E(B-V)=0.79$ (Winkler
1997\nocite{1997MNRAS.287..481W}). It is therefore likely
associated with, or located further than RCW\,158. A second rather
extincted O-type star (\object{ALS\,4657}) is located at $\rm \sim18$\,pc
distance. No young clusters reside within a projected distance of
65\,pc.



{\it Our result:} A physical cluster is not detected from the density maps.

%% file: STARS_TEX/hd169515.tex
\subsection{HD\,169515 ($\equiv$ RY\,Sct)}

RY\,Sct is a well-known double-line massive eclipsing binary. The object is
surrounded by a young compact nebula ($<2''$) with an unusual geometry, including a
concentric set of ionized rings (Smith et al. 1999\nocite{1999AJ....118..960S}).
HD\,169515 is located in a field containing a number of dark clouds, masers and
H\,{\sc ii} regions.  The high-mass protostellar candidate \object{IRAS\,18223-1243} is
located at +5 arcminutes east. Therefore the field shows evidence for active star
formation. The cluster \object{NGC\,6604} is a young object with an age of
4\,Myr and a distance of 1.1\,kpc in \object{Sharpless 2-54} 
(Battinelli et al. 1994\nocite{1994A&AS..104..379B}). The cluster's distance to
the Sun is comparable to that of HD\,169515.


{\it Our result:} The high resolution density map in Fig.\,\ref{f_hd1695} does not
reveal a cluster near HD\,169515.  The 2MASS map in the bottom right panel has a
rather structured stellar density distribution due to large scale extinction on the
west side of the map. Therefore the density "peak" on the 2MASS map near HD\,169515
is not unambiguously identified as a physical cluster.

\begin{table}[h]
 {
 \begin{center}
  \begin{tabular*}{0.49\textwidth}
   {@{\extracolsep{\fill}}lccccc}
\hline
\hline
 Cluster Name & Ang Dist & Lin Dist & Age  & Dist & Ref \\
	      & (deg)    & (pc)     & (Myr)& (kpc) & \\ 
\hline
NGC\,6604 & 1.9 &65 & 4 & $1.1\pm0.3$ & 1 \\
\hline
 \end{tabular*}
 \end{center}
\vspace{-0.2cm}
 {\tiny 1: Battinelli et al.(1994)\nocite{1994A&AS..104..379B}}
 }
\end{table}


%% file: FIG_TEX/Fig15AP.tex
\begin{figure*}[ht]
 \center
 \includegraphics[height=15cm,width=16cm,angle=-90]{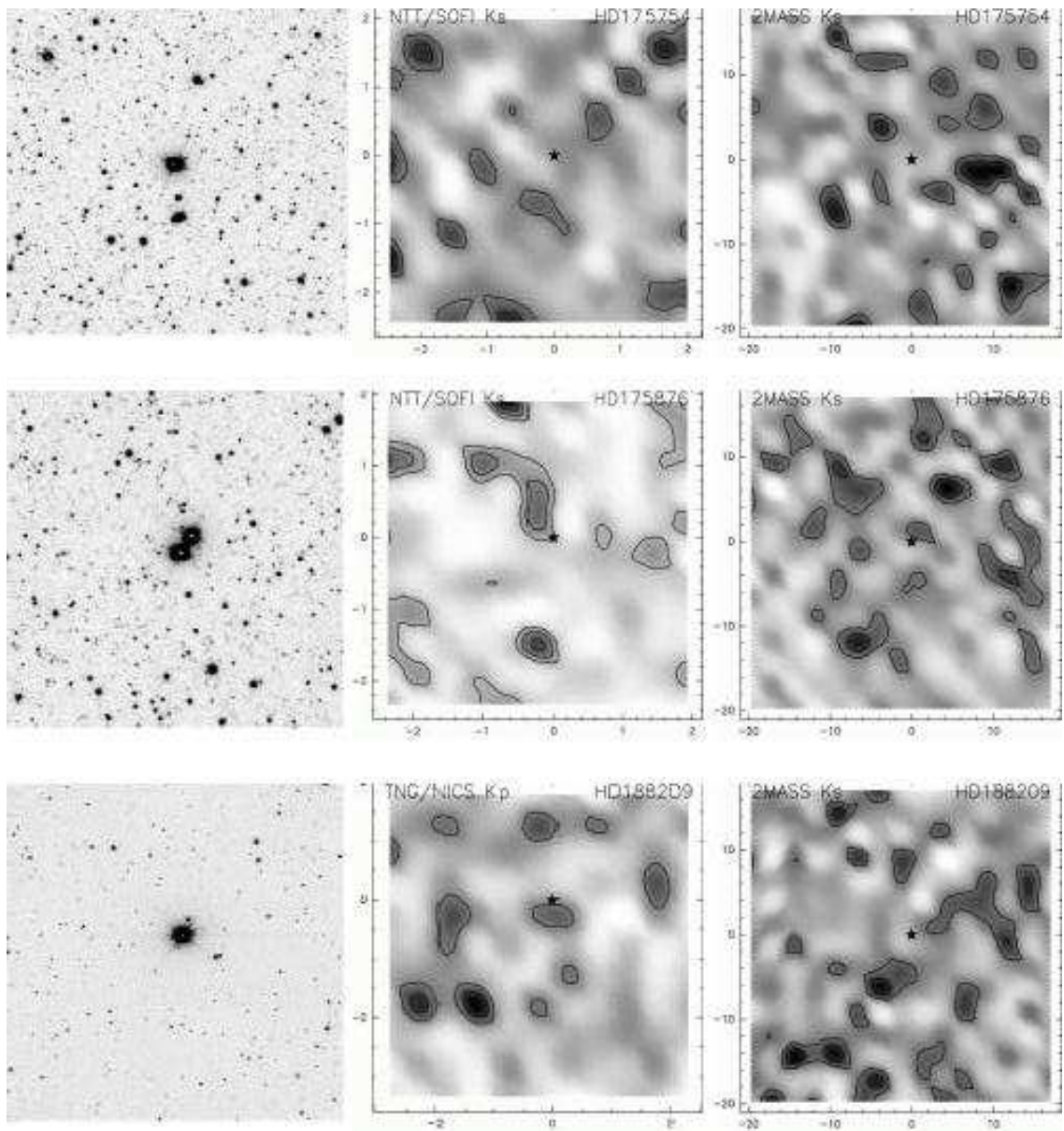}
\caption[]{As in Fig.\,\ref{f_hd3687} for stars HD\,175754, HD\,175876, and HD\,188209.}
\label{f_hd1882}
\end{figure*}

%% file: STARS_TEX/hd175754.tex
\subsection{HD\,175754 and HD\,175876}
\label{hd1757}

HD\,175754 and HD\,175876 are described together. Their
similarities led Walborn \& Fitzpatrick (2000\nocite{2000PASP..112...50W}) to suggest
that they are physically related.

HD\,175754 is an emission line star and a single object, both
optically and spectroscopically. It is located $\rm 400\,pc$ below the
Galactic plane. HD\,175876 is an optical binary from the Lindroos
sample (Lindroos 1985\nocite{1985A&AS...60..183L}), located at
$\rm z\simeq-420\,pc$. The two O-type stars have an angular separation of 1.3
degrees, that translates to $\sim50$~pc.  Both objects could be
related to the 'blowout' observed in the Scutum Supershell
(Callaway et al. 2000\nocite{2000ApJ...532..943C}).

{\it Our result:} The density maps do not reveal clusters near HD\,175754 and HD\,175876.








%% file: STARS_TEX/hd175876.tex
\subsection{HD\,175876}
See Sect.\,\ref{hd1757}

%% file: STARS_TEX/hd188209.tex
\subsection{HD\,188209}

The object has a relatively large distance from the Galactic plane ($\rm
z\simeq-330\,pc$) and is classified as a candidate runaway star by Stone
(1979)\nocite{1979ApJ...232..520S} on the basis of a large ($\rm >40km\,s^{-1}$)
peculiar space velocity. A number of studies have searched for a binary nature of
HD\,188209, which is suggested by its intricate line-profile variations (e.g.
Fullerton et al. 1996)\nocite{1996ApJS..103..475F}. Recently taken echelle spectra
detect a 6.4 days
period (Israelian et al.  2000\nocite{2000MNRAS.316..407I}), however the authors
choose not to infer a binary nature from these measurements. HD\,188209 is
associated with the X-ray source \object{1RXS\,J195159.2+470133}. No young clusters within
65\,pc from HD\,188209 are known. Closest early type star is the B0.5\,III star
\object{HD\,188439} at 28\,pc distance. 


{\it Our result:} The surface density maps do not reveal a cluster
near the star.

%% file: FIG_TEX/Fig16AP.tex
\begin{figure*}[ht]
 \center
 \includegraphics[height=15cm,width=16cm,angle=-90]{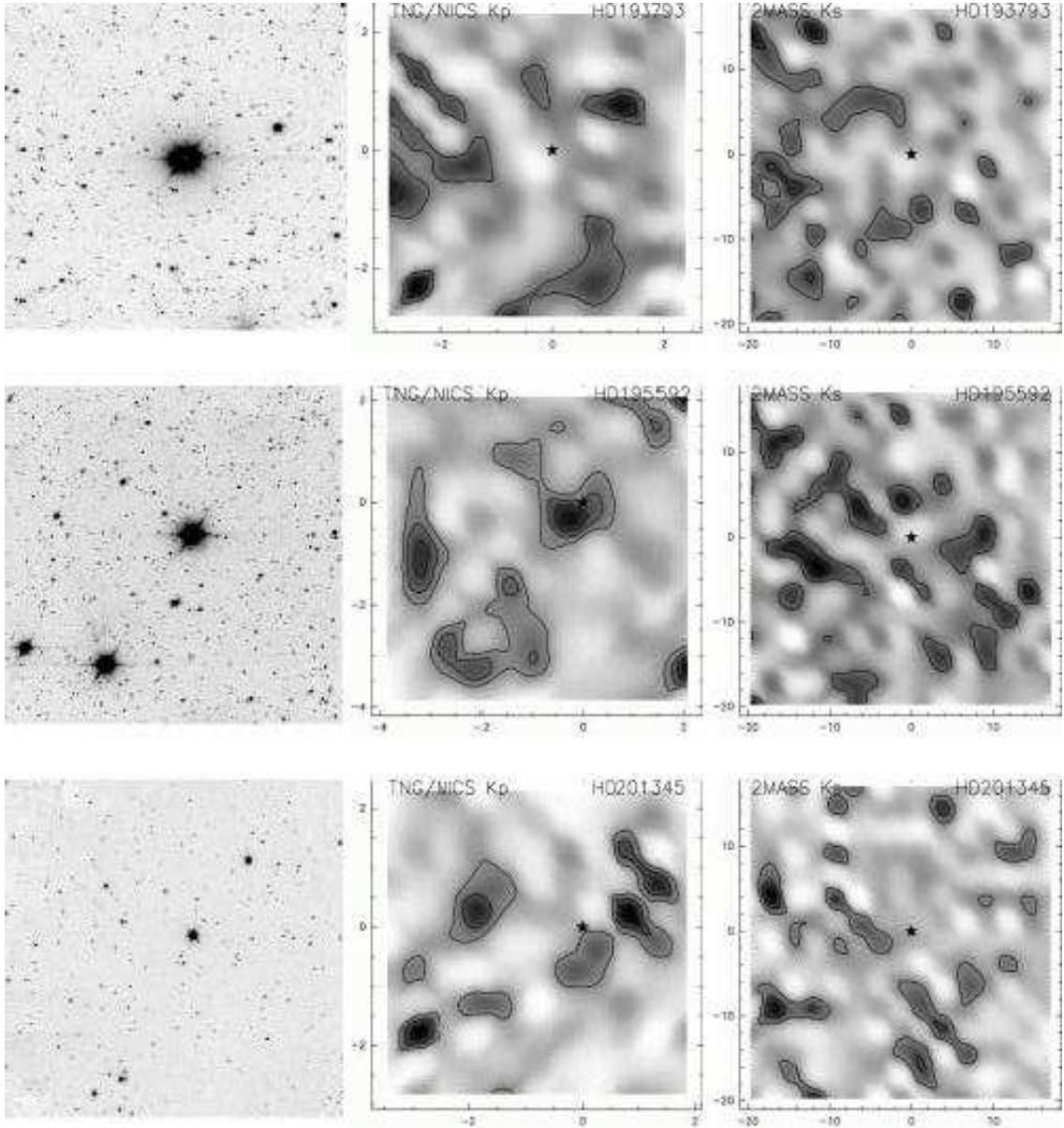}
\caption[]{As in Fig.\,\ref{f_hd3687} for stars HD\,193793, HD\,195592, and HD\,201345.}
\label{f_hd2013}
\end{figure*}

%% file: STARS_TEX/hd193793.tex
\subsection{HD\,193793}
The object is a triple system consisting of a 6 arcsec visual binary and the
spectroscopic WR-O binary WR\,140. A system prototypical for colliding wind
phenomena (Monnier et al.  2002\nocite{2002ApJ...567L.137M}), it shows
periodical dust formation when close to periastron (e.g. Williams
1996\nocite{1996RMxAC...5...47W}). Optically, the field of HD\,193793 shows
clear evidence for wisps of ionized gas. IRAS $60\,\mu$m imaging shows the field
in the interior of a ring-like IR emission structure.

{\it Our result:} The density maps do not reveal a cluster near
this star.

%% file: STARS_TEX/hd195592.tex
\subsection{HD\,195592}

HD\,195592 is an emission line star which is located in the Cygnus
X region. The region contains swirls of interstellar ionized gas
(see Dickel et al. 1969)\nocite{1969A&A.....1..270D} and shows the
presence of dust extinction. Noriega-Crespo et al. (1997)
identified an IRAS bow shock near this star.
We mention two young clusters in the mini-table that, although without age
estimates, are likely to be very young as they are thought to be
the driving source of massive outflows (see Itoh et al 
1999\nocite{1999MNRAS.304..406I}; Hodapp
1994\nocite{1994ApJS...94..615H}). However both SF region seem to be
located further out into the Galaxy.

{\it Our result:} The high resolution density map from TNG imaging
in Fig.\,\ref{f_hd2013} indicates the presence of a cluster near
HD\,195592. This cluster is however not detected on the 2MASS
density map.


\begin{table}[h]
 {
 \begin{center}
  \begin{tabular*}{0.49\textwidth}
   {@{\extracolsep{\fill}}lccccc}
\hline
\hline
 Cluster Name & Ang Dist & Lin Dist & Age & Dist & Ref \\
	& (deg) & (pc) & (Myr) & (kpc) & \\
\hline
 \object{W 75N IR Cluster}           & 2.2 &54  & - & 2.0 & 1\\
 \object{DR 21}                   &  2.5 &61  & - & 3.0 & 2\\
\hline
 \end{tabular*}
 \end{center}
\vspace{-0.2cm}
 {\tiny 1: Hodapp 1994\nocite{1994ApJS...94..615H}; 2: Itoh et al. 1999\nocite{1999MNRAS.304..406I}}
 }
\end{table}

%% file: STARS_TEX/hd201345.tex
\subsection{HD\,201345}

HD\,201345 is located at a distance of $\rm z\simeq-300\,pc$ below the Galactic
plane and has a relatively high peculiar radial velocity of
$\rm 29.5\,km\,s^{-1}$ reported by Gies and Bolton
(1986\nocite{1986ApJS...61..419G}). The star's photosphere shows
evidence for enhanced nitrogen abundance (Walborn
1976\nocite{1976ApJ...205..419W}). There are no clusters nor early
type ($\rm <B\,5$) stars known within 65\,pc of the object for our
adopted distance of 1.9\,kpc.


{\it Our result:} We find no indication of a cluster from the density maps near the star.

%% file: FIG_TEX/Fig17AP.tex
\begin{figure*}
 \center
 \includegraphics[height=15cm,width=5cm,angle=-90]{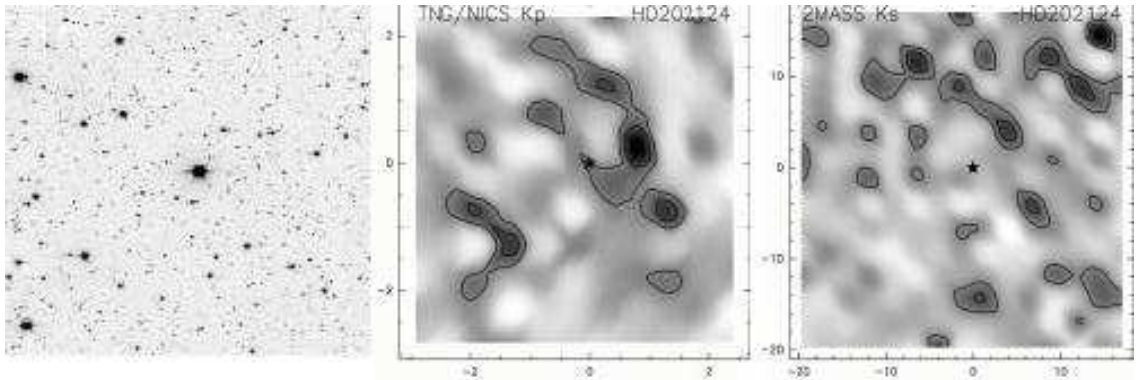}
\caption[]{As in Fig.\,\ref{f_hd3687} for star HD\,202124.}
\label{f_hd2021}
\end{figure*}

%% file: STARS_TEX/hd202124.tex
\subsection{HD\,202124}

Spectroscopic measurements show that HD\,202124 is likely to be a single object. It
is located in the direction of the Cygnus superbubble, a structure related to the
OB associations located along the Orion local spiral arm (see Uyaniker et al.
2001\nocite{2001A&A...371..675U}).  Despite this, there are no young cluster known
within 65\,pc from HD\,202124 . An early B2 type star is found at $\sim 30$\,pc
(HD\,202253). 

{\it Our result:} The stellar density maps do not reveal any cluster near the
target star.
